\documentclass{article}
\usepackage[toc]{appendix}
\usepackage{graphicx} 
\usepackage{hyperref}
\hypersetup{colorlinks=true, linkcolor=black, citecolor=blue, urlcolor=blue}
\usepackage[style=numeric-comp,sorting=none]{biblatex}
\usepackage{amsmath}
\usepackage{amssymb}
\usepackage{amsfonts} 
\numberwithin{equation}{section}
\usepackage{abstract}
\usepackage{subfig}
\usepackage{authblk}
\usepackage{imakeidx}
\usepackage{tabularx}
\usepackage{adjustbox}
\usepackage{geometry}
\title{\textbf{Tails from the Bulk: Gravitational Decay in AdS$_5$}}

\author{John~R.~V.~Crump}
\author{Jorge~E.~Santos}

\affil{\it{Department of Applied Mathematics and Theoretical Physics,\\\it{University of Cambridge, Wilberforce Road, Cambridge, CB3 0WA, UK}}}

\date{}

\begin{document}

\maketitle

\begin{abstract}


We study gravitational perturbations of Schwarzschild–AdS black holes in $d = 5$ and identify a regime of late-time power-law decay for smooth initial data. Based on an analysis of the quasinormal mode spectrum, we predict and characterise this decay behaviour. We perform fully nonlinear numerical evolutions with long integration times that support the prediction and exhibit no signs of instability. Remarkably, the decay is modulated by a universal oscillatory pattern, consistent with subleading corrections from a large-angular-momentum (eikonal) analysis of the quasinormal mode spectrum.
    
\end{abstract}

\tableofcontents

\section{Introduction}
The stability of black hole solutions within general relativity is an old but still active topic of research. While the linear mode stability of the Schwarzschild was established long ago \cite{Regge:1957td,Vishveshwara:1970cc,Zerilli:1970se,Price:1971fb,Bardeen:1973xb,Moncrief:1974am,10.1063/1.524181}, the full linear stability of the Schwarzschild black hole was only proven in 2016 in the seminal work of Dafermos, Holzegel, and Rodnianski \cite{Dafermos:2016uzj}.

Building on the roadmap laid out in \cite{Dafermos:2016uzj}, the nonlinear stability of the Schwarzschild black hole was subsequently established in \cite{Dafermos:2021cbw} (see also \cite{Klainerman:2017nrb} for results concerning more restricted classes of initial data).

One might wonder whether any stability results exist for spacetimes with different asymptotics. Indeed, this is the case for Schwarzschild–de Sitter. In particular, Kerr–de Sitter black holes have been shown to be nonlinearly stable in \cite{Hintz:2016gwb}, and this result also applies to the Schwarzschild–de Sitter case as a special limit. This is perhaps not a surprising result, given that Schwarzschild itself is stable. Indeed, one might expect the expansion induced by a positive cosmological constant to help perturbations decay.

The case of a negative cosmological constant—\emph{i.e.}, Anti-de Sitter (AdS) asymptotics—is much less well understood. The strongest motivation for studying such spacetimes stems from the so-called AdS/CFT correspondence \cite{Maldacena:1997re,Gubser:1998bc,Witten:1998qj,Aharony:1999ti}, which, in an appropriate limit, relates states of certain strongly coupled quantum field theories with a large number of degrees of freedom—defined on a fixed background geometry—to classical solutions of specific supergravity theories. In its original form, the classical solutions are those of type IIB supergravity with AdS$_5 \times S^5$ asymptotics. In particular, via a Freund–Rubin ansatz \cite{Freund:1980xh}, any classical solution of Einstein’s theory with a negative cosmological constant can be uplifted to a classical solution of type IIB supergravity with the desired asymptotics. It is therefore paramount not only to understand the moduli space of stationary solutions in such a theory but also to analyze their stability properties.

In the AdS/CFT context, black holes play a pivotal role. They are dual to thermal states in the quantum field theory \cite{Witten:1998qj} and often exhibit interesting universal properties. However, black holes with AdS asymptotics can exhibit behaviour markedly different from that of asymptotically flat black holes. For instance, black resonators—black holes with a single Killing vector field—exist with AdS asymptotics \cite{Dias:2015rxy,Ishii:2018oms}; multiple black hole solutions can coexist at fixed charges \cite{Hartnoll:2008kx,Dias:2010ma,Dias:2011tj,Dias:2015rxy}, and a plethora of near-horizon geometries exist with no residual spatial symmetries \cite{Horowitz:2022leb}, to name but a few.

Given the vast number of possible black hole solutions with AdS asymptotics, it is paramount to study their stability properties, as only stable solutions can truly represent genuine equilibrium states. As a first step, one might consider studying the stability of the ground state—namely, pure AdS. Indeed, Dafermos and Holzegel conjectured a nonlinear instability in which the reflecting boundary of AdS allows small but finite-energy perturbations to grow and eventually collapse into a black hole \cite{Dafermos2006,DafermosHolzegel2006}. This contrasts with Minkowski and de Sitter spacetimes, for which nonlinear stability has long been established \cite{Christodoulou:1993uv,friedrich86}.

The first numerical indication supporting the instability of AdS was presented in the groundbreaking work by Bizon and Rostworowski \cite{Bizon:2011gg}. Since then, this subject has garnered significant interest from both numerical and theoretical perspectives \cite{Dias:2011ss,Dias:2012tq,Buchel:2012uh,Buchel:2013uba,Maliborski:2013jca,Bizon:2013xha,Maliborski:2012gx,Maliborski:2013ula,Baier:2013gsa,Jalmuzna:2013rwa,Basu:2012gg,2012arXiv1212.1907G,Fodor:2013lza,Friedrich:2014raa,Bizon2014,Maliborski:2014rma,Abajo-Arrastia:2014fma,Balasubramanian:2014cja,Bizon:2014bya,Balasubramanian:2015uua,daSilva:2014zva,Craps:2014vaa,Basu:2014sia,Deppe:2014oua,Dimitrakopoulos:2014ada,Horowitz:2014hja,Buchel:2014xwa,Craps:2014jwa,Basu:2015efa,Yang:2015jha,Fodor:2015eia,Okawa:2015xma,Bizon:2015pfa,Dimitrakopoulos:2015pwa,Green:2015dsa,Deppe:2015qsa,Craps:2015iia,Craps:2015xya,Evnin:2015gma,Menon:2015oda,Jalmuzna:2015hoa,Evnin:2015wyi,Freivogel:2015wib,Dias:2016ewl,Evnin:2016mjx,Deppe:2016gur,Dimitrakopoulos:2016tss,Dimitrakopoulos:2016euh,Rostworowski:2016isb,Bantilan:2017kok,Jalmuzna:2017mpa,Rostworowski:2017tcx,Martinon:2017uyo,Moschidis:2017lcr,Moschidis:2017llu,Dias:2017tjg,Choptuik:2017cyd,Choptuik:2018ptp}. Notably, this instability has been rigorously proven for the spherically symmetric, pressureless Einstein–massless Vlasov system \cite{Moschidis:2017lcr,Moschidis:2017llu}.

The origin of the instability has been attributed to two key features of AdS: its fully resonant spectrum \cite{Bizon:2011gg,Dias:2011ss}, and its non-dispersive nature \cite{Dafermos2006,DafermosHolzegel2006}. In other words, perturbative modes not only persist without decaying, but the frequencies in the spectrum also obey nontrivial linear relations, enabling energy transfer between modes that can ultimately result in black hole formation. One may then wonder whether this instability mechanism persists in the presence of a black hole. In \cite{Dias:2012tq}, it was argued that asymptotically anti-de Sitter solutions with a non-resonant spectrum should be nonlinearly stable (if linearly stable). In particular, the claim was that an asymptotically resonant spectrum alone is generally insufficient to trigger a nonlinear instability of the type first observed in \cite{Bizon:2011gg}. The class of spacetimes considered in \cite{Dias:2012tq} includes Schwarzschild-AdS and Kerr-AdS black holes below the Hawking-Reall bound \cite{Hawking:1999dp}.

This conclusion stands in contrast to the conjecture put forward in \cite{Holzegel:2011uu}, as well as more recent numerical results in \cite{Figueras:2023ihz}. The latter identifies an intermediate regime characterized by $1/\log t$ decay, consistent with the analysis in \cite{Holzegel:2011uu}, and also presents evidence for a nonlinear instability affecting both Schwarzschild-AdS and Kerr-AdS black holes. In \cite{Holzegel:2011uu}, the inverse-logarithmic decay was attributed to the presence of stable trapping, which inhibits the decay of high-$\ell$ modes.

Stable trapping leaves a distinctive imprint on the quasinormal mode (QNM) spectrum of Schwarzschild-AdS black holes. To describe this, let us recall a few facts about linear perturbations around Schwarzschild-AdS backgrounds. Because the Schwarzschild-AdS black hole is stationary, its domain of outer communications admits a timelike Killing vector field $\partial/\partial t$, allowing perturbations to be decomposed into Fourier modes $e^{-i \omega t}$. The spherical symmetry of the background also permits decomposition in terms of scalar, vector, or tensor spherical harmonics labeled by an angular number $\ell$.

The QNM spectrum of Schwarzschild-AdS black holes \cite{Horowitz:1999jd,Cardoso:2001bb,Michalogiorgakis:2006jc,Dias:2011ss,Dias:2012tq} consists of mode solutions that are smooth across the future event horizon, normalizable at infinity, and have finite energy. Due to stable trapping, it is known that the QNM spectrum contains modes whose imaginary parts approach zero exponentially fast with increasing $\ell$, i.e., ${\rm Im}(\omega L) \sim -e^{-\alpha \ell}$ for some constant $\alpha > 0$, where $L$ is the AdS length scale. These slowly decaying modes become asymptotically resonant \cite{Dias:2011ss}, and thus may serve as seeds for a nonlinear instability.

One might then wonder whether \cite{Dias:2012tq} makes an implicit assumption that may not hold for generic initial data. Indeed, \cite{Dias:2012tq} does not discuss the most generic \emph{radial overtones} configurations, which might appear to be a relatively benign assumption, but could in fact be a source of instability. In that work, it is argued that if the spectrum QNMs of a given solution in AdS$_d$ is asymptotically resonant, taking the form
\begin{equation}
\omega_{\ell\,p} = \omega^{\rm AdS}_{\ell\,p} + K_d\,\ell^{-\frac{d-3}{2}} + \ldots,
\label{eq:res}
\end{equation}
where $d$ is the bulk spacetime dimension, $p$ the radial overtone number, and $K_d$ a constant depending on $p$, the energy and angular momentum of the background spacetime, and the type of perturbation (scalar, vector, or tensor), then initial data in $H^s$ with $s > (d+1)/2$ should be immune to a nonlinear instability\footnote{Here we consider $4 \leq d \leq 6$, since for $d \geq 7$, stability instead requires $s > (3d - 5)/4$ \cite{Dias:2012tq}, and the analysis in \cite{Dias:2012tq} does not exclude the possibility of a nonlinear instability for sufficiently rough data when $d \geq 8$.}. This result is particularly appealing, as initial data in $H^{\frac{d+1}{2} + \epsilon}$, with $\epsilon > 0$, is believed to correspond to the lowest regularity for which Minkowski spacetime remains nonlinearly stable \cite{DafermosRodnianskiPC}.

However, Eq.~(\ref{eq:res}) was derived under the assumption\footnote{J.~E.~Santos would like to acknowledge a conversation with Christoph Kehle in which this point was raised.} that the initial data includes only modes with $p \ll \ell$. It is therefore natural to ask what happens when $p$ grows with $\ell$. We note, however, that this must be done with care, since if $p$ grows too fast, the natural norm to consider may no longer be a Sobolev norm, but rather something more akin to a BV norm. In this regime, it is conceivable that the leading correction term of order $\ell^{-\frac{d-3}{2}}$ could vanish. If so, the analysis in \cite{Dias:2012tq} would then suggest that a nonlinear instability may arise for sufficiently rough initial data. Note that the same analysis also suggests that \emph{smooth initial data} should not lead to a nonlinear instability. Moreover, the expected decay for such data is not of the form $1/\log t$, but rather follows a power-law in $t$, possibly modulated by subleading corrections (as discussed in Section 2.3). It is therefore surprising that \cite{Figueras:2023ihz} reports observing a $1/\log t$ decay even for smooth initial data (see Eq.~(3.19) in \cite{Figueras:2023ihz}). We will focus our analysis on the case $d = 5$ and present strong evidence against the inverse $\log t$ decay, at least for initial data that preserves the $SO(3)$ symmetry (which ensures the background's zero angular momentum is maintained). Although our longest evolution encompasses more than 4,000 bounces off the AdS wall, we cannot rule out the possibility of a longer timescale after which an instability sets in. Nevertheless, the fact that our analytic prediction closely matches the late-time evolution lends credibility to the expected power-law decay for generic (smooth) perturbations predicted in \cite{Dias:2012tq}. Along the way, we uncovered a plethora of new phenomena related to the unexpected behavior of overtones, which we detail in the subsequent sections.

This paper is organized as follows. In section \ref{sec:setup}, we describe the setup of the problem using ingoing Bondi-Sachs coordinates and study the perturbative expansion of the metric near the AdS boundary. We then derive a prediction for the late time behaviour, taking the form of a power law, using properties of the quasinormal modes of Schwarzschild-AdS$_5$ and assuming analytic initial data. We discuss the timescales involved in observing the predicted behaviour and then lay out the marching orders for the equations of motion. In section \ref{sec:results}, results are presented for two numerical case studies for a small black hole and a large black hole. In section \ref{sec:discussion}, we close with a discussion of our findings and possible directions for future work. We supply appendices with details of our numerical methods.

\section{Setup of the problem}\label{sec:setup}
\subsection{Ingoing Bondi-Sachs coordinates}

The bulk 5-dimensional Einstein-Hilbert action with a negative cosmological constant $\Lambda = -\frac{6}{L^2}$ is given by

\begin{equation}
    S = \frac{1}{16 \pi G_{5}} \int \mathrm{d}^5 x \sqrt{-g} \left( 
R + \frac{12}{L^2} \right) \label{eqn:bulkaction}
\end{equation}

\noindent where $L$ is the AdS length. Varying this action with respect to the metric produces the equation of motion (EOM)

\begin{equation}
    G_{a b} \equiv R_{a b} + \frac{4}{L^2} g_{a b} = 0\label{eqn:Einstein}.
\end{equation}

\noindent We use ingoing Bondi-Sachs coordinates so that the Einstein equation takes the form of nested first-order ODEs along ingoing null characteristics \cite{Balasubramanian_2014, bondi1960}. The line element takes the form

\begin{equation}
    \mathrm{d}s^2 = \frac{1}{z^2} \left\{ B^2 \left( -V \mathrm{d}v^2 - 2 \mathrm{d}v \mathrm{d}z \right) + e^{2 \chi} \left[ \frac{1}{A^4} \left( \frac{\mathrm{d}x}{\sqrt{1-x^2}} - \sqrt{1-x^2} U \mathrm{d}v \right)^2 + (1-x^2) A^2 \mathrm{d}\Omega_2^2 \right] \right\}\label{eqn:lineelement}
\end{equation}

\noindent where we have fixed $L=1$ and $\mathrm{d}\Omega_2^2$ is the line element on the unit radius round two-sphere. The transformation of $x\in[-1,1]$ by $x=\cos\theta$ brings us back to the usual polar coordinate $\theta \in [0,\pi]$ on $S^3$.

There is a timelike conformal boundary at $z=0$, and surfaces of constant $v$ are surfaces of ingoing null characteristics moving away from the boundary in the radial $z$-direction. With these coordinates we can recover vacuum AdS$_5$ with $\chi = U = 0$, $A=B=1$, $V=1+z^2$. We can also obtain the Schwarzschild-AdS$_5$ solution with 

\begin{equation}
    U=0, \qquad A=B=1, \qquad \chi=\log{(y_{+})}, \qquad V=(1-z^2)\left[1+z^2\left(1+\frac{1}{y_{+}^2}\right)\right]\label{eqn:boundaryconditions}
\end{equation}

\noindent where the black hole horizon is located at $z=1$ with radius $y_{+}$ in units of the $\mathrm{AdS}$ length. It is around this solution that we will add perturbations.

We restrict the metric components to depend only on the time coordinate $v$, radial coordinate $z$, and polar coordinate $x$, so that, along with the form of the line element \eqref{eqn:lineelement}, we preserve an $SO(3)$ of the $SO(4)$ symmetry of the $S^3$. This effectively reduces the system from $4+1$ to a $2+1$ problem. These coordinates have a remaining gauge freedom which we can fix by choosing $\chi$ to take the form

\begin{equation}
    \chi(v,z,x) = \log \left(y_{+} + z^4 \chi_4(v,x)\right)\label{eqn:gaugefix}.
\end{equation}

\noindent It is simpler to implement the numerics when the coordinate domain has a fixed size, so we use \eqref{eqn:gaugefix} to place the apparent horizon at $z=1$ and excise the region $z>1$. This is possible because the Einstein equation \eqref{eqn:Einstein} is hyperbolic, so information from behind the apparent horizon cannot reach the computational domain $z\in[0,1]$. Setting $z=1$ to be an apparent horizon means requiring the expansion of outgoing null rays to vanish on this hypersurface. This condition can be written as

\begin{equation}
    \Theta \equiv \left[ \partial_v \chi + \frac{1}{2 z} \left( 1 - z \partial_z \chi \right) V + (1-x^2) U \partial_x \chi + \frac{1}{3} \left((1-x^2) \partial_x U - 3 x U \right) \right] \bigg|_{z=1} = 0\label{eqn:horizoncondition}.
\end{equation}

\noindent These coordinates, together with initial data prescribed on the hypersurface $v=0$, boundary conditions at $z=0$, and the apparent horizon condition \eqref{eqn:horizoncondition}, give us a well-defined Cauchy problem.

\subsection{Boundary expansion and the CFT stress tensor}

The AdS/CFT correspondence tells us that results from the CFT can be obtained from the metric and its derivatives evaluated at the conformal boundary $z=0$. To make this connection, it is useful to write each metric component as a power series in $z$ and solve the EOM \eqref{eqn:Einstein} order by order in $z$. This is also useful for the numerics as we can then define new variables that subtract off the leading order terms. This avoids the need to take high-order derivatives numerically when solving the EOM and when isolating CFT quantities at the boundary, as well as improving numerical stability.

The leading order terms are determined by our choice of boundary conditions and correspond to source terms in the CFT. We impose Dirichlet boundary conditions so that the metric approaches Schwarzschild-AdS$_5$ \eqref{eqn:boundaryconditions} at leading order. The following boundary expansion emerges

\begin{align}
    A(v,z,x) &= 1 + (1-x^2) A_4 (v,x) z^4 + (1-x^2) A_5 (v,x) z^5 + \mathcal{O}(z^6) \label{eqn:Aexpansion}\\
    B(v,z,x) &= 1 - \frac{3}{2 y_{+}} \chi_{4}(v,x) z^4 - \left[ \frac{9}{8 y_{+}^2} \chi_4 (v,x)^2 + 2 (1-x^2)^2 A_4 (v,x)^2\right] z^8 + \mathcal{O}(z^9) \\
    U(v,z,x) &= \frac{1}{y_{+}^2} U_4 (v,x) z^4 + \mathcal{O}(z^5) \\
    V(v,z,x) &= 1 + \frac{z^2}{y_{+}^2} + \left[ -\left(1 + \frac{1}{y_{+}^2}\right) + \frac{5}{y_{+}} \chi_4 (v,x) + \frac{1}{y_{+}^2} V_4 (v,x) \right] z^4 + \mathcal{O}(z^5)\label{eqn:Vexpansion}.
\end{align}

\noindent The boundary expansion of the EOM also gives us the equations for the time evolution of $U_4$, $V_4$, and $A_4$ which are

\begin{align}
    \partial_v U_4 &= \frac{1}{4 y_{+}^2} \partial_x V_4 + 4 \left( (1-x^2) \partial_x A_4 - 5 x A_4 \right)\label{eqn:DvU4}\\
    \partial_v V_4 &= \frac{4}{3} \left( (1-x^2) \partial_x U_4 - 3 x U_4 \right)\label{eqn:DvV4} \\
    \partial_v A_4 &= \frac{1}{15 y_{+}^2} \partial_x U_{4} + A_5.
\end{align}

\noindent The AdS/CFT dictionary tells us how the boundary expansion of the bulk metric is related to the boundary CFT stress tensor \cite{Skenderis_2006}. Following the procedure of \cite{Balasubramanian_1999} we add to the action \eqref{eqn:bulkaction} the Gibbons-Hawking-York boundary term to yield a well-defined variational problem \cite{Hawking_1996}, as well as boundary terms built from local curvature invariants to cancel out divergences arising from the diverging Weyl factor of the metric as $z\to 0$.

We interpret the variation of these boundary terms with respect to the boundary metric, evaluated on-shell with respect to the bulk EOM, as the vacuum expectation value of the CFT stress tensor $\langle T_{\mu \nu} \rangle$. This procedure yields

\begin{equation}
    \langle T_{\mu \nu} \rangle = \frac{1}{8 \pi G_{5}} \left( K_{\mu \nu} - \gamma_{\mu \nu} K - 3 \gamma_{\mu \nu} + \frac{1}{2} E_{\mu \nu} \right)
\end{equation}

\noindent where $\gamma_{\mu \nu}$ is the boundary metric on a surface of constant $z$, $K_{\mu \nu}$ is the extrinsic curvature of the surface, and $E_{\mu \nu}$ is the Einstein tensor associated to $\gamma_{\mu \nu}$. Plugging in our boundary expansion, the non-vanishing components are

\begin{align}
    8 \pi G_5 \langle T_{v v} \rangle &= \frac{1}{y_{+}^4} \left( \frac{3}{8} (1 + 2 y_{+}^2)^2 - \frac{3}{2} y_{+}^2 V_{4} \right) z^2 + \mathcal{O}(z^3) \label{eqn:Tvv}\\
    8 \pi G_5 \langle T_{v x} \rangle &= -2 \, U_4 \, z^2 + \mathcal{O}(z^3) \\
    8 \pi G_5 \langle T_{x x} \rangle &= \frac{1}{(1-x^2)}\frac{1}{y_{+}^2}\left( \frac{1}{8}(1+2y_{+}^2)^2 -\frac{1}{2}y_{+}^2 V_{4} - 8 y_{+}^4 (1-x^2) A_{4} \right) z^2 + \mathcal{O}(z^3) \\
    8 \pi G_5 \langle T_{\Omega_2 \Omega_2} \rangle &= (1-x^2)\frac{g_{\Omega_2 \Omega_2}}{y_{+}^2}\left( \frac{1}{8}(1+2y_{+}^2)^2 -\frac{1}{2}y_{+}^2 V_{4} + 4 y_{+}^4 (1-x^2) A_{4} \right) z^2 + \mathcal{O}(z^3)
\end{align}

\noindent where $g_{\Omega_2 \Omega_2}$ is the 2-sphere metric. Being derived from the variation of the boundary action with respect to $\gamma_{\mu\nu}$ this satisfies the conservation equation $D_{\mu}\langle T^{\mu\nu}\rangle=0$. This can be verified explicitly using the EOM \eqref{eqn:DvU4} and \eqref{eqn:DvV4}. The limit as $y_{+}\to 0$ appears singular, but this is rectified by changing coordinates to boundary time $t=\frac{v}{y_{+}}$ and radius $r=\frac{y_{+}}{z}$.

\subsection{Prediction of late time behaviour}

Perturbing a Schwarzschild-AdS black hole, we expect an initial period of nonlinear behaviour as radiation bounces off the conformal boundary and falls into the horizon, followed by a ringdown period where the decaying perturbation becomes small with respect to the black hole length scale $y_{+}$. The behaviour of the remaining small perturbation can be studied by approximating the dynamics using the linearised Einstein equation \cite{HolzegelI}. Solutions can be written as sums of quasinormal modes (QNMs) \cite{Cardoso_2001}. Whilst QNMs do not form a complete basis for the solution space \cite{Warnick_2014}, we assume that they capture the dominant features of the late time behaviour. 

QNMs oscillate and decay with characteristic frequencies and are labelled by an angular quantum number $\ell$, two azimuthal numbers for the $S^2$, and an overtone number $n$ \cite{Cardoso_2003}. The preserved $SO(3)$ symmetry of our problem means that we consider only perturbations with the azimuthal numbers set to zero. We ignore higher overtones $n\geq1$ for each $\ell$ as these decay faster than the fundamental $n=0$ modes. We write the QNM Ansatz for a linearised perturbation of the boundary variables $A_4$, $\chi_4$, $U_4$, and $V_4$ as

\begin{align}
    A_4(v,x) &= \sum_{\ell=2}^{\infty} \tilde{A}_{4 \, \ell}(v) \partial_x \partial_x Y_{\ell}(x) & 
    \chi_4 (v,x) &= \sum_{\ell=0}^{\infty} \tilde{\chi}_{4 \, \ell}(v) Y_{\ell}(x) \\
    U_4 (v,x) &= \sum_{\ell=1}^{\infty} \tilde{U}_{4 \, \ell}(v) \partial_x Y_{\ell}(x) &
    V_4 (v,x) &= \sum_{\ell=0}^{\infty} \tilde{V}_{4 \, \ell}(v)  Y_{\ell}(x)
\end{align}

\noindent where $Y_{\ell}(x)$ are Chebyshev polynomials of the second kind, which are eigenfunctions of the Laplacian on $S^3$. The $v$-dependences of the mode amplitudes $\tilde{A}_{4 \, \ell}(v)$, $\tilde{\chi}_{4 \, \ell}(v)$, $\tilde{U}_{4 \, \ell}(v)$, and $\tilde{V}_{4 \, \ell}(v)$ can be written as the real parts of complex exponentials, each one taking the form

\begin{equation}
    \tilde{f}_{\ell}(v) = \operatorname{Re} \left( \tilde{f}_{\ell}(v_0) e^{-i \omega_\ell (v-v_0)} \right) \label{eqn:QNMmodebehaviour}
\end{equation}

\noindent where the mode frequencies $\omega_\ell$ are complex. The real part $\operatorname{Re}\left(\omega_\ell\right)$ is the mode oscillation frequency, and the imaginary part $\operatorname{Im}\left(\omega_\ell\right)$, when $\operatorname{Im}\left(\omega_\ell\right)<0$, is the decay rate. A positive imaginary part, $\operatorname{Im}\left(\omega_\ell\right)>0$, would indicate a linear instability, however no such modes exist for Schwarzschild-AdS$_5$ \footnote{However, linearly unstable modes do exist in the related problem of small Schwarzschild-AdS$_5\times S^5$ black holes in type IIB supergravity \cite{Hubeny_2002, Buchel_2015}.}.

The $U_4$ and $V_4$ EOM \eqref{eqn:DvU4}-\eqref{eqn:DvV4} give us the frequencies $\omega_{\ell=0}=0$ and $\omega_{\ell=1}=\frac{1}{y_{+}}$. Because these frequencies have no imaginary part, the $\ell=0$ and $\ell=1$ modes do not decay. These correspond to the conserved quantities $\tilde{V}_{4 \, \ell=0}$ and $\tilde{V}_{4 \, \ell=1}^2+16 y_{+}^2 \tilde{U}_{4 \, \ell=1}^2$, which can also be derived from the conservation of the boundary stress tensor. Modes with $\ell>1$ have $\operatorname{Im}(\omega_\ell)<0$.

Each mode is kept away from the horizon by a potential barrier due to its angular momentum. The size of the potential barrier increases with $\ell$, allowing high $\ell$ modes to have very long lifetimes as they slowly leak across the barrier and fall into the black hole. For $\ell\gg 1$ the decay rates approach degenerating exponential behaviour \cite{HolzegelII}, asymptotically behaving as

\begin{equation}
    \operatorname{Im}(\omega_\ell) \sim -e^{-C \ell+\kappa} \label{eqn:decayrate}
\end{equation}

\noindent where $C=C(y_{+})>0$, and $\kappa =\kappa (y_{+})\in \mathbb{R}$. WKB analysis \cite{Dias_2012} can be used to show that the dependence of $C$ on the horizon radius $y_{+}$ can be written approximately as

\begin{equation}
    C(y_{+}) = \frac{2}{1+2 y_{+}^2} \left(\sqrt{1+y_{+}^2} \operatorname{arcsch}\,y_{+} - y_{+} \operatorname{arccot}\,y_{+} \right)\,. \label{eqn:WKBslope}
\end{equation}

\noindent To study the decay rate of the solution, we track the norm of $V_4 (v,x)$ on $S^3$ as $V_4 (v,x)$ is a physically meaningful quantity corresponding to the perturbation of the energy density of the boundary CFT \eqref{eqn:Tvv}. We define the norm

\begin{equation}
    \| V_{4} \|^2 (v) = \frac{2}{\pi} \int_{-1}^{1} \sqrt{1-x^2}\, V_4 (v,x)^2 \mathrm{d}x = \sum_{\ell=0}^{\infty} \tilde{V}_{4 \, \ell}(v)^2 \label{eqn:norm}
\end{equation}

\noindent where the sum over the modes $\tilde{V}_{4 \, \ell}$ follows from the orthogonality of $Y_{\ell}$. 

Choosing initial data such that $\tilde{V}_{4 \, \ell=0}(v)=\tilde{V}_{4 \, \ell=1}(v)=\tilde{U}_{4 \, \ell=1}(v)=0$, we expect that at late times the spectrum will be dominated by a long tail of large $\ell$ modes after the faster decaying low $\ell$ modes have attenuated. The norm at late times $v>v_0$ can then be approximated by

\begin{equation}
    \| V_{4} \|^2 (v) \approx \sum_{\ell=2}^{\infty} \tilde{V}_{4 \, \ell}(v_0)^2 \exp{\left( -2 e^{-C\ell+\kappa} (v-v_0) \right)} \label{eqn:longtail}
\end{equation}

\noindent where we have used the QNM $v$-dependence \eqref{eqn:QNMmodebehaviour} and large $\ell$ decay rate \eqref{eqn:decayrate}, whilst suppressing the contributions from $\operatorname{Re}(\omega_\ell)$.

We consider initial data ${V}_{4}(v_0, x)$ which is analytic in $x$. Analyticity implies that the mode amplitudes degenerate at least exponentially in the mode number, $\lvert \tilde{V}_{4 \, \ell}(v_0) \rvert \leq \exp{\left(-\alpha \ell+\beta\right)}$ for some $\alpha, \beta \in \mathbb{R}$, $\alpha>0$. With initial data that has a spectrum of this form, and further approximating the sum over the modes of the long tail in \eqref{eqn:longtail} as an integral over $\ell$, we obtain

\begin{equation}
    \| V_{4} \|^2 (v) \approx \int_{2}^{\infty} \exp{\left(-2\alpha \ell+2\beta\right)} \exp{\left( -2 e^{-C\ell+\kappa} (v-v_0) \right)} \, \mathrm{d}\ell.
\end{equation}

\noindent Making the substitution $u = 2 e^{-C\ell+\kappa} (v-v_0)$ brings this to the form

\begin{equation}
    \| V_{4} \|^2 (v) \approx \left( v-v_0 \right)^{-\frac{2\alpha}{C}} \left[ \frac{1}{C}\exp{\left( 2\beta - \frac{2\alpha}{C}(\kappa + \log{2})\right)} \int_{0}^{2(v-v_0)\exp{(-2C + \kappa)}} u^{\frac{2\alpha}{C}-1} e^{-u}  \mathrm{d}u  \right]. \label{eqn:integralapprox}
\end{equation}

\noindent The integral contribution tends to $\Gamma(\frac{2\alpha}{C})$ exponentially fast in $v$, so at late times $v\gg v_0$ we are left with a power law decay of the form

\begin{equation}
    \| V_{4} \|^2 (v) \propto \, v^{-\frac{2\alpha}{C}}. \label{eqn:prediction}
\end{equation}

\noindent To investigate subleading terms, taking a log of \eqref{eqn:longtail} and dropping $v_0$ we have

\begin{equation}
    \log{(\| V_{4} \|^2 (v))} = \log{\left(\sum_{\ell=2}^{\infty}  \exp{\left(-2\alpha \ell+2\beta  -2 v e^{-C\ell+\kappa}\right)}  \right)} \approx \max_{l} \left(-2\alpha \ell+2\beta  -2 v e^{-C\ell+\kappa}\right). \label{eqn:LSE approx}
\end{equation}

\noindent Approximating $\ell$ as continuous, the $\ell$ that maximises the expression at time $v$ is $\ell_{\rm max} = -\frac{1}{C} \left( \log{\left(\frac{\alpha}{Cv}\right)} - \kappa \right)$, yielding

\begin{equation}
    \| V_{4} \|^2 (v) \approx v^{-\frac{2\alpha}{C}} \Big( \frac{2\alpha}{C}\Big)^{\frac{2\alpha}{C}} e^{-\frac{2\alpha}{C}} \exp{\Big( 2 \beta - \frac{2\alpha}{C} (\kappa + \log{2}) \Big)},
\end{equation}

\noindent which recovers the previous approximation \eqref{eqn:integralapprox} with Stirling's approximation to $\Gamma(\frac{2\alpha}{C})$. Factoring this out of \eqref{eqn:LSE approx}, we have

\begin{equation}
    \| V_{4} \|^2 (v) = v^{-\frac{2\alpha}{C}} e^{2 \beta} \sum_{\ell=2}^{\infty} \exp{\left( \frac{2\alpha}{C}\left(\log{(v)-C \ell}\right) -2 e^{\left( \log{(v)}-C \ell\right)+\kappa} \right)}. \label{eqn:Poisson form}
\end{equation}

\noindent The exponent $\frac{2\alpha}{C}\left(\log{(v)-C \ell}\right) -2 e^{\left( \log{(v)}-C \ell\right)+\kappa}$ has a maximum at $\ell=\ell_{\rm max}$ and decays linearly with $\ell$ for $\ell > \ell_{\rm max}$, and decays exponentially with $\ell$ for $\ell<\ell_{\rm max}$. This means that at late times the sum is well-approximated by replacing the lower bound of summation $\ell=2$ with $\ell=-\infty$. The sum is then periodic in $\log{(v)}$ with period $C$ and can be evaluated with the Poisson summation formula. Letting $\log{(v)}=C t$, this yields

\begin{equation}
    \| V_{4} \|^2 (v) \approx v^{-\frac{2\alpha}{C}} e^{2 \beta} \sum_{n=-\infty}^{\infty} e^{2\pi i n t} \int_{-\infty}^{\infty} \exp{\left( (2\alpha-2\pi i n) t'-2 e^{Ct'+\kappa} \right)} dt'.
\end{equation}

\noindent The substitution $u=2 e^{Ct'+\kappa}$ brings this to the form

\begin{equation}
    \| V_{4} \|^2 (v) \approx v^{-\frac{2\alpha}{C}} \frac{1}{C} \exp{\left( 2\beta - \frac{2\alpha}{C}(\kappa + \log{2})\right)} \sum_{n=-\infty}^{\infty} \exp{\left( \frac{2\pi i n}{C} \big( \log{(v)} +\kappa+\log{2}\big) \right)} \Gamma\Big( \frac{2\alpha}{C}-\frac{2\pi i n}{C} \Big).
\end{equation}

\noindent The series can be truncated at small $|n|$ as $|\Gamma\big( \frac{2\alpha}{C}-\frac{2\pi i n}{C} \big)|$ decays rapidly with $|n|$. Truncating at $n=0$ recovers \eqref{eqn:integralapprox}. Truncating at $|n|=1$ reveals the first subleading term

\begin{equation}
    \| V_{4} \|^2 (v) \approx v^{-\frac{2\alpha}{C}} \frac{1}{C} \exp{\left( 2\beta - \frac{2\alpha}{C}(\kappa + \log{2})\right)} \left( \Gamma\Big(\frac{2\alpha}{C}\Big) +2 |\Gamma\Big( \frac{2\alpha}{C}-\frac{2\pi i }{C} \Big)|\cos{\Big(\frac{2\pi}{C} \log{(v)} + \phi \Big)} \right) \label{eqn:prediction subleading}
\end{equation}

\noindent where the phase shift is $\phi=\frac{2\pi}{C}(\kappa+\log{2})+\mathrm{arg}(\Gamma( \frac{2\alpha}{C}-\frac{2\pi i }{C}))$. Therefore at late times we expect to observe power law decay in $v$ with exponent $-\frac{2\alpha}{C}$ and subleading oscillations in $\log{(v)}$ with period $C(y_{+})$.

To observe this effect in the numerical solutions, it is useful to write this as the gradient of a log-log plot. This becomes

\begin{equation}
    \frac{\mathrm{d} \log{\| V_{4} \|^2 (v)}}{\mathrm{d} \log{(v)}} \approx -\frac{2 \alpha}{C} - \frac{2\pi}{C} \frac{\sigma \sin{(\frac{2\pi}{C} \log{(v)}+\phi)}}{1+\sigma \cos{((\frac{2\pi}{C} \log{(v)}+\phi))}} \label{eqn:loglog grad pred}
\end{equation}

\noindent where $\sigma=2|\Gamma\big( \frac{2\alpha}{C}-\frac{2\pi i }{C} \big)| \Gamma \big( \frac{2\alpha}{C}\big)^{-1}$. The amplitude of the oscillatory term in \eqref{eqn:loglog grad pred} is $\frac{2\pi}{C} \sigma (1-\sigma^2)^{-\frac{1}{2}}$.

An assumption in this prediction is that the tail of the spectrum contains all modes up to $\ell=\infty$ but in numerical solutions we only have access to a finite number of modes. However, we still expect to observe the predicted behaviour as from \eqref{eqn:Poisson form} we see that the dominant contributions to the power law decay at time $v$ come from the peak of the spectrum at $v$ and the immediately adjacent modes.

Whilst the norm $||V_4||^2$ is physically meaningful, it is limited as $V_4$ is defined at the AdS timelike boundary. To capture global behaviour of the spacetime, we define the norm  $||I_1||^2$ given by

\begin{equation}
    ||I_1||^2 (v) = \int_{\Sigma_v} \mathrm{d}^4 x \sqrt{-g} \, \left(C_{abcd} C^{abcd}\right)^2 \label{eqn:I1 def}
\end{equation}

\noindent where $C_{abcd}$ is the Weyl tensor and $\Sigma_v$ is the characteristic null hypersurface at time $v$ stretching from the AdS boundary at $z=0$ to the apparent horizon at $z=1$. Whilst an apparent horizon is not a gauge invariant notion, in practice it rapidly coincides with the event horizon during the time evolution. In this spacetime, Ricci decomposition relates the curvature scalar $I_1=C_{abcd}C^{abcd}$ to the Riemann tensor by $I_1 = R_{abcd}R^{abcd}-40$. On the exact Schwarzschild-AdS$_5$ solution \eqref{eqn:boundaryconditions} this is given by $I_1=72 (1+\frac{1}{y_{+}^2})^2 z^8$. Consequently, to study the decay of perturbations we track the quantity $||I_1||^2(v) - 864 \pi^2 y_{+}^3 ( 1+\frac{1}{y_{+}^2} )^4$ which, assuming solutions settle down to Schwarzschild-AdS$_5$ at late times, should decay to zero. Repeating the analysis applied to $||V_4||^2$ for $||I_1||^2$, with the same assumptions, produces the same late time power law prediction.

\subsection{How late is late?} \label{sec:how late}

The prediction of a late time power law \eqref{eqn:prediction} is based on the assumption that QNMs dominate the late time dynamics. However, this does not mean that we expect to see this behaviour as soon as the early time nonlinear regime has ended and the QNM regime has begun. The other assumption in making the prediction is that the QNM decay rates -$\operatorname{Im}(\omega_\ell)$ degenerate exponentially in $\ell$ \eqref{eqn:decayrate}. This is only true asymptotically for large $\ell$, so we may expect an intermediate regime where the solution behaves as a sum of QNMs but where the spectrum is dominated by lower $\ell$ modes with decay rates that are not well-described by the WKB result. The predicted late time regime may then begin to appear when the lower $\ell$ modes have decayed to the point where the peak of the spectrum lies at sufficiently high $\ell$.

\begin{figure}[h]
    \centering
    \includegraphics[width=1.0\linewidth]{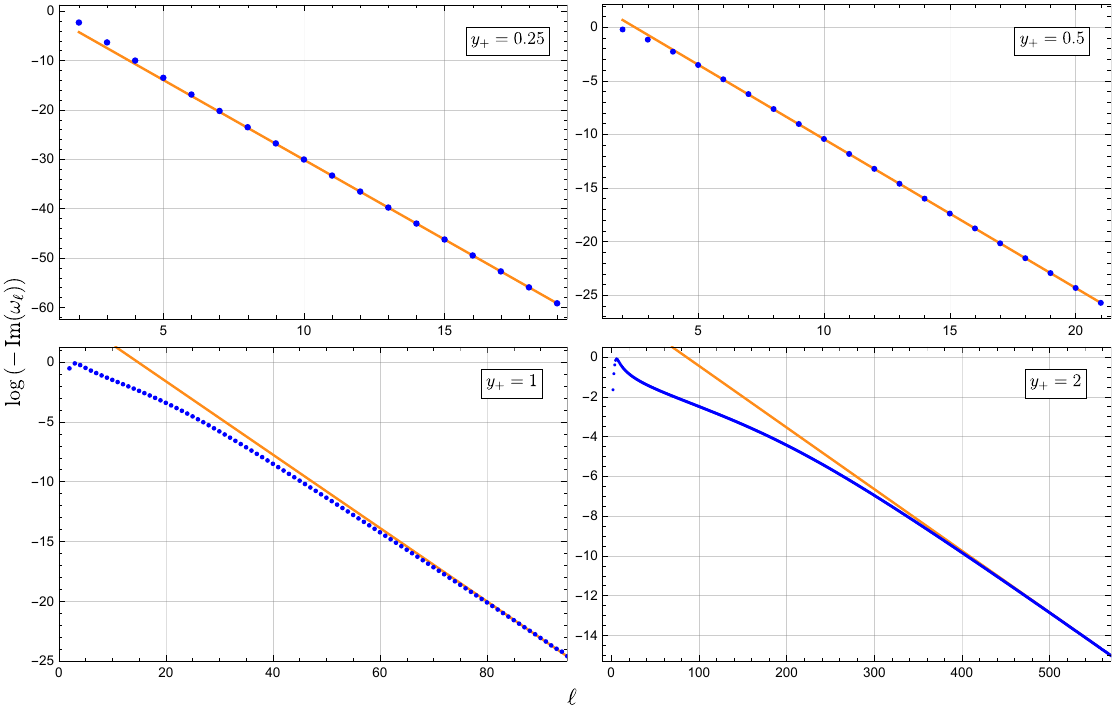}
    \caption{Plots of the log of the quasinormal fundamental mode decay rates $\log{(-\operatorname{Im}(\omega_\ell))}$ for horizon radii $y_{+}\in \{0.25,0.5,1,2\}$ are shown in blue. Linear fits with gradients given by the WKB result $-C(y_{+})$ are shown in orange.}
    \label{fig:decayrates}
\end{figure}

For example, the decay rates for $y_{+}=1$, shown in Fig. \ref{fig:decayrates}, indicate that the WKB result is a good approximation only for $\ell \gtrsim 40$. This effect becomes more noticeable as $y_{+}$ increases, illustrated in Fig. \ref{fig:decayrates} which shows the decay rates for $y_{+}=2$ where the WKB result is a good approximation only for $\ell \gtrsim 300$. The approximation is better beginning at lower $\ell$ for smaller black holes as the perturbation of the metric away from pure AdS, due to the presence of the black hole, at radii around the angular momentum potential barrier becomes smaller. 

\begin{figure}[h]
    \centering
    \includegraphics[width=1.0\linewidth]{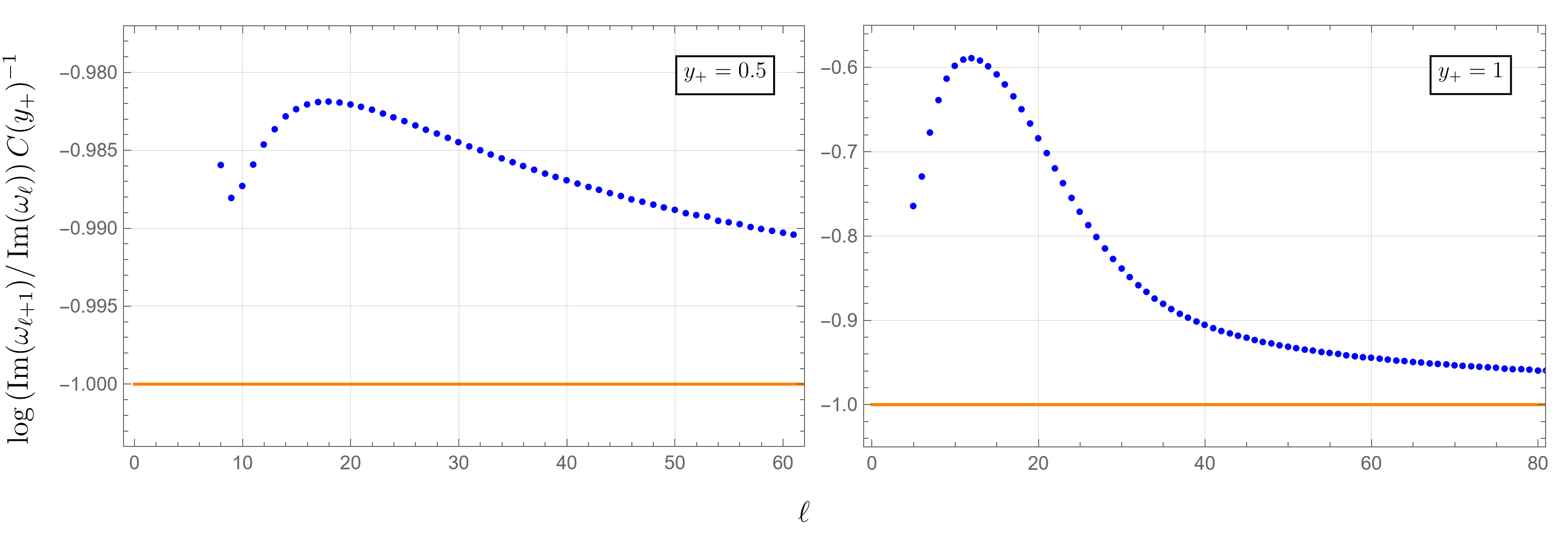}
    \caption{Plots of $\log{(\operatorname{Im}(\omega_{\ell+1})/\operatorname{Im}(\omega_{\ell})})\,C(y_{+})^{-1}$ for $y_{+}=0.5$ and $y_{+}=1$. WKB analysis shows that this tends to $-1$ at large $\ell$. The rate at which it approaches $-1$ depends on $y_{+}$.}
    \label{fig:decay rate ratios}
\end{figure}

The power law prediction $ \| V_{4} \|^2 (v) \propto \, v^{-\frac{2\alpha}{C}}$ depends on $C(y_{+})$ where $-C(y_{+})$ is the asymptotic value of $\log{(\operatorname{Im}(\omega_{\ell+1})/\operatorname{Im}(\omega_{\ell})})$ as $\ell\to\infty$. Fig. \ref{fig:decay rate ratios} shows the approach to this asymptote for $y_{+}=0.5$ and $y_{+}=1$. We see that this approach can be slow, with the difference from the asymptote for $y_{+}=0.5$ becoming less than 1\% only after $\ell=58$. The decay rate of the $\ell=58$ mode for $y_{+}=0.5$ is $-\operatorname{Im}(\omega_{\ell=58})=2.58 \times 10^{-34}$. This indicates a typically long evolution time until the peak of the spectrum reaches this mode and the observed power law exponent differs from the predicted value by less than 1\%.

Approximating the $v$-dependence of each mode as $\tilde{V}_{4 \, \ell}(v) = \exp{(-\alpha \ell +\beta)\exp{(-v \,e^{-C\ell+\kappa})}}$, then the time $v_{\ell}$ at which mode $\ell$ begins to dominate is given by $\log{(v_{l})} = C \ell-\kappa+\log{(\alpha)}-\log{(e^{C}-1)}$. This indicates that we may need to wait for exponentially long evolution times before we begin to observe the predicted behaviour. This can be mitigated by choosing initial data with small $\alpha$. We may also want the $\ell$ at which the WKB result becomes a good approximation to be small by choosing a small black hole radius $y_{+}$, however this is hindered by $C(y_{+})$ becoming large. 

The prediction assumes that the late time behaviour is dominated by the fundamental $n=0$ mode at each $\ell$ and ignores the contribution of higher $n$ overtones. At high $\ell$, the overtones also have long lifetimes with decay rates that degenerate exponentially with $\ell$ so may still contribute significantly to the tail of the spectrum at late times. The mode number $\ell$ of the peak of the contribution to the spectrum from the $n=0$ modes increases with $C^{-1} \log{(v)}$, as do the peaks of the contributions from each higher $n$. However, the peak of the spectrum of each higher overtone will remain at a fixed number of modes higher than the peak of the $n=0$ spectrum. Since the predicted power law decay is predominantly sourced by the modes close to the peak of the spectrum rather than the tail, we expect that the overtones will not contribute significantly to the power law behaviour.

\subsection{Marching orders and new variables}

Initial data is specified by choosing $A$, $\chi_4$, $U_4$, and $V_4$ at time $v=0$. The apparent horizon condition must be satisfied on the initial slice, so in practice we specify $A$, $U_4$, and $V_4$ and then find $\chi_4$ such that \eqref{eqn:horizoncondition} is satisfied. Further details about specifying $\chi_4$ on the initial slice are given in \autoref{appendix:AH}. We can then use the EOM \eqref{eqn:Einstein} to solve for the remaining metric components on the time slice and their time derivatives. Firstly, $G_{zz}=0$ can be used to solve for $B$. To find $U$, we first define the intermediate variable $\Pi$

\begin{equation}
    \Pi = \frac{e^{5 \chi}}{z^3 A^4 B^2} \partial_z U.\label{eqn:defPi}
\end{equation}

\noindent $\Pi$ can be found using $G_{zx}=0$, and then the definition of $\Pi$ \eqref{eqn:defPi} can be used to obtain $U$. We define the intermediate variables $d_t \chi$ and $d_t A$ by

\begin{align}
    d_{t} \chi &= \partial_v \chi - \frac{1}{2} \left( V -1 -\frac{z^2}{y_{+}^2} \right) \left( \partial_z \chi - \frac{1}{z}\right) \label{eqn:defdtX} \\
    d_{t} A &= \partial_v A - \frac{1}{2} V \partial_z A\label{eqn:defdtA}.
\end{align}

\noindent Using $g^{ab}G_{ab}=0$ with $a,b \in \{x, \Omega_2\} $ we can obtain $d_t \chi$, and then with $G_{xx}=0$ we can get $d_t A$. These equations are all first-order ODEs which must be integrated in $z$ and require boundary conditions. We use Dirichlet boundary conditions at $z=0$, which can be found using the boundary expansions \eqref{eqn:Aexpansion}-\eqref{eqn:Vexpansion}. 

The combination $G^{z}_{\,v}+(1-x^2)U G^{z}_{\,x}=0$ at $z=1$, with $G_{zz}=0$ and the apparent horizon condition used to eliminate terms proportional to $V^2$, $\partial_v V$, $\partial_z V$ and $\partial_v B$, is a second-order elliptic equation that can be solved for the horizon value of $V$. This requires two boundary conditions which amount to requiring that $V$ be an $S^3$ scalar, which we impose at the poles $x=\pm 1$. The apparent horizon condition \eqref{eqn:horizoncondition} can then be rearranged to find $\partial_v \chi_4$. The definition of $d_t \chi$ \eqref{eqn:defdtX} can be rearranged to give us $V$ everywhere in the bulk, which allows us to obtain $\partial_v A$ from the definition of $d_t A$ \eqref{eqn:defdtA}. Knowing $\partial_v A$ and $\partial_v \chi_4$, along with $\partial_v U_4$ and $\partial_v V_4$ from \eqref{eqn:DvU4} and \eqref{eqn:DvV4}, we can integrate $A$, $\chi_4$, $U_4$, and $V_4$ forwards in $v$ to the next time slice.

We use the boundary expansion to define new variables to use for the numerics. This allows us to cancel out algebraically rather than numerically the terms in the EOM which diverge as $z \to 0$ arising from the leading order terms in the boundary expansion. We use the new variables $Q_1$, $q_1$, $\ldots$ , $q_6$ defined by

\begin{align}
    A &= 1 + (1-x^2) \, Q_1 \, z^4 \\
    B &= 1 - \frac{3}{2 y_{+}} \chi_4 \, z^4 + q_1 \, z^8 \\
    \Pi &= y^3 \, q_2 \\
    U &= \frac{1}{y_{+}^2}  \,q_3  \, z^4 \\
    d_t \chi &= y_{+}^3\, e^{-3\chi} \left( -\frac{1}{2} \left(1+\frac{1}{y_{+}^2}\right) + \frac{1}{y_{+}^2} q_4 \right) \,z^3 \\
    d_t A &= (1-x^2) \, A \,e^{-\frac{3}{2}\,\chi} \, q_5 \,z^3 \\
    V &= (1-z^2)\left( 1+ z^2 \left( 1+\frac{1}{y_{+}^2} \right) \right) + \frac{1}{y_{+}^2}\, q_6 \, z^4.
\end{align}

\noindent The Dirichlet boundary conditions we impose at $z=0$ in terms of these new variables take the form

\begin{align}
    q_1(v,0,x) &= -\frac{9}{8 y_{+}^2} \chi_4(v,x)^2 -2 (1-x^2)^2 Q_1(v,0,x)^2 \label{eqn:Dirichlet BCs first}\\
    q_2(v,0,x) &= 4 U_4(v,x) \\
    q_3(v,0,x) &= U_4(v,x) \\
    q_4(v,0,x) &= \frac{5 y_{+}}{2} \chi_4(v,x) + \frac{1}{2} V_4(v,x) \\
    q_5(v,0,x) &= -2 y^{\frac{3}{2}}Q_1(v,0,x). \label{eqn:Dirichlet BCs last}
\end{align}

\noindent In the numerical implementation, we solve for $p_i(v,z,x)=q_i(v,z,x)-q_i(v,0,x)$ and then add on the boundary terms to recover $q_i$ as we find that this leads to better conditioned matrices.

To find solutions numerically, we discretise the spacetime with a pseudo-spectral method. The polar $x$-direction is gridded with Chebyshev-Gauss-Lobatto collocation points \cite{Trefethen, CanutoI, Berrut2004BarycentricLI} and a spectral element discontinuous Galerkin (DG) method is used in the radial $z$-direction \cite{CanutoII}. The elements are gridded with Gauss-Legendre-Lobatto points and the element interfaces are managed by a Lax-Friedrichs flux. We employ a fourth-order Runge-Kutta method (RK4) in the $v$-direction. Further details of the numerical method are supplied in \autoref{appendix:numerics}.

\section{Results} \label{sec:results}

\subsection{\texorpdfstring{A small black hole: $y_{+}=0.5$}{A small black hole: y+ = 0.5}}

We investigate the case of a small black hole with $y_{+}=0.5$. We choose initial data which is analytic on $x\in [-1,1]$ given by $V_4(0,x)=10^{-2}\left((1.3-x)^{-1} - \frac{1}{5}(13-\sqrt{69}) - \frac{2}{25} (119-13 \sqrt{69})\,x \right)$, $U_4(0,x)=0$, $Q_{1}(0,z,x)= 10^{-4} \,z \, e^{z} \cos{(x-\frac{1}{2})}$. The subtracted terms in $V_4(0,x)$ ensure that the non-decaying modes $\ell=0,1$ are initially zero. The horizon condition \eqref{eqn:horizoncondition} is used to determine $\chi_4 (0,x)$.

\begin{figure}[p]
    \centering
    \includegraphics[width=1.0\linewidth]{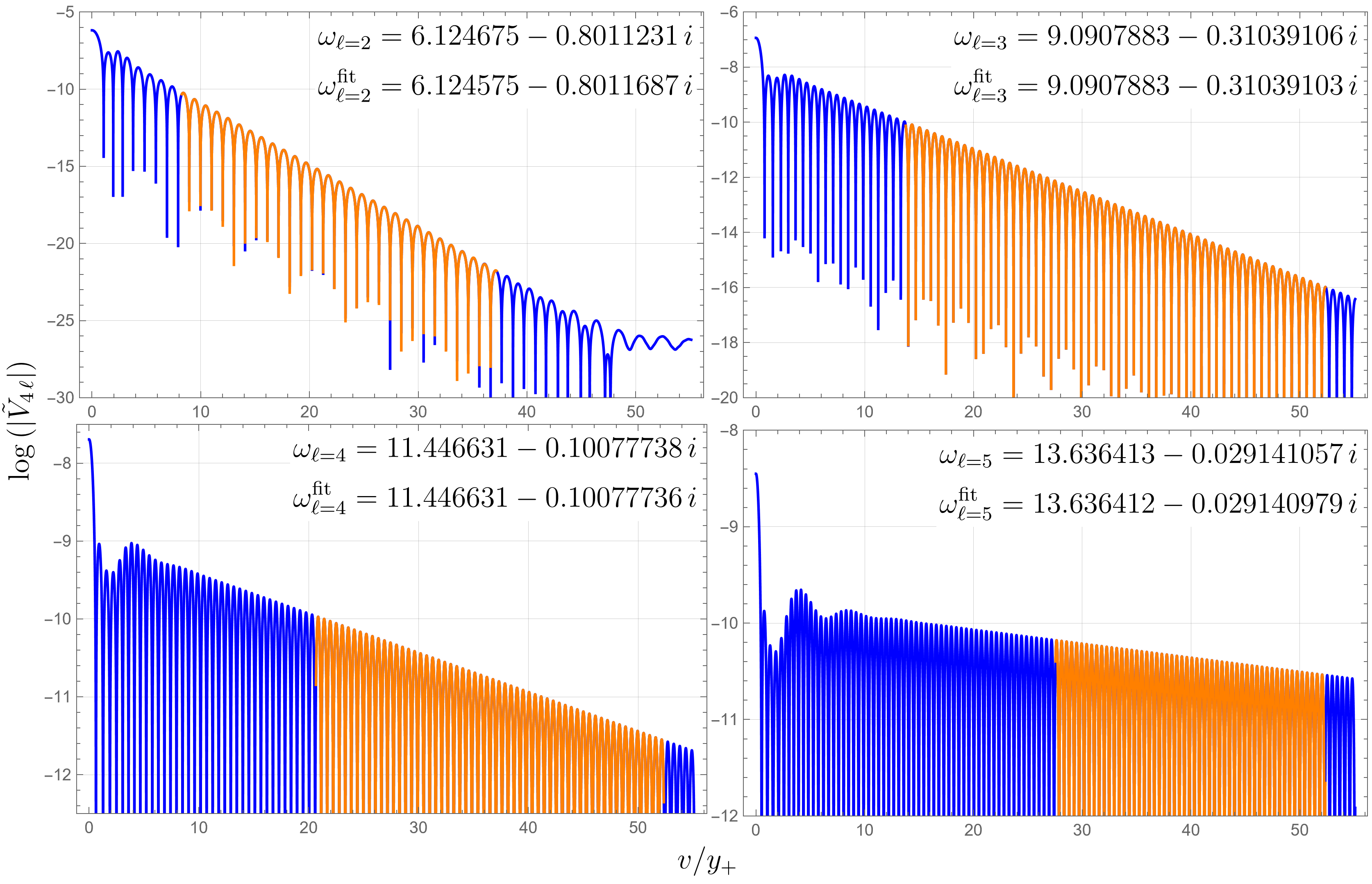}
    \caption{Plots of the log of the $\ell=2,3,4,5$ modes of $V_4$ at early times are shown in blue. Fits to a single QNM are shown in orange. The fitted frequencies $\omega_l^{\rm{fit}}$ are shown alongside the frequencies $\omega_l$ of the fundamental $n=0$ modes of the linearised equations. QNM behaviour ceases when a mode decays to the point where nonlinear interactions of other modes takes over, seen here for $\ell=2$ after $v/y_{+} \gtrsim 50$.}
    \label{fig:QNM plots}
\end{figure}

\begin{figure}[p]
\begin{tabular}{c}
    \includegraphics[width=0.97\linewidth]{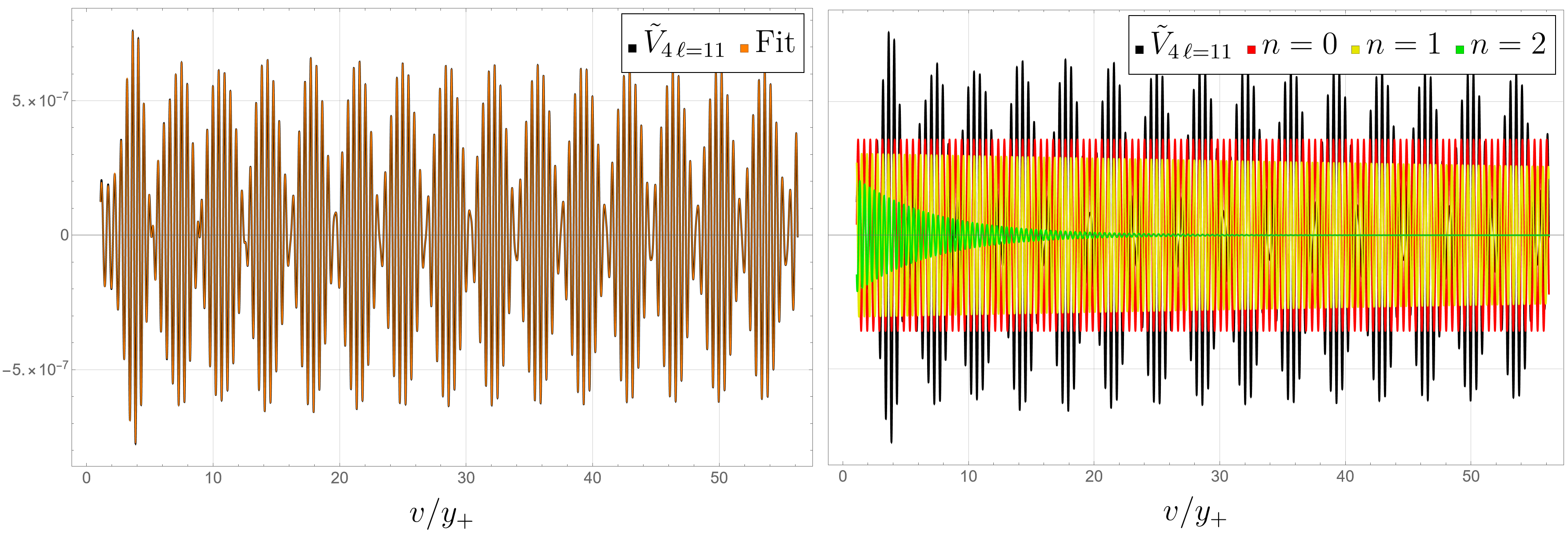} \\
    \includegraphics[width=0.6\linewidth]{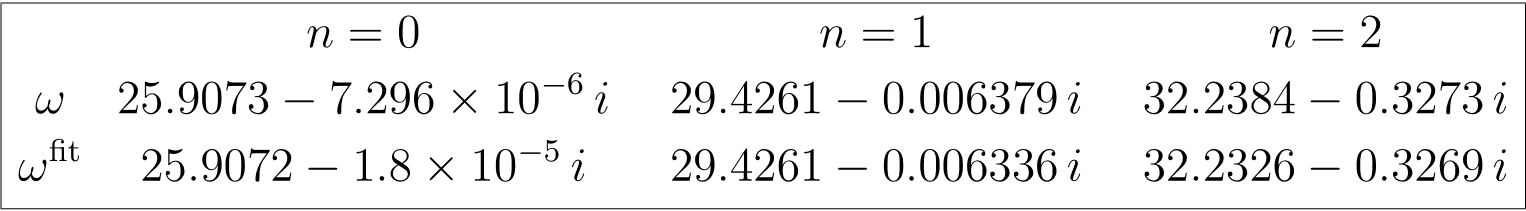}
\end{tabular}
    \caption{\textit{(Left)} Plot of the $\ell=11$ mode of $V_4$ at early times and a fit to a sum of three QNMs which we identify as the $n=0,1,2$ overtones. Interference of the $n=0$ and $n=1$ overtones produces prominent low-frequency beating. \textit{(Right)} Plot with overlays showing the fitted overtones. The fitted frequencies $\omega^{\mathrm{fit}}$ are shown alongside the frequencies $\omega$ from the linearised equations. The decay rate for $n=0$ has a large error due to the time interval being short but it can be recovered accurately over the full evolution.}
    \label{fig:l=11 QNM plots}
\end{figure}

\begin{figure}[h]
\begin{tabular}{cc}
     \includegraphics[width=0.49\linewidth]{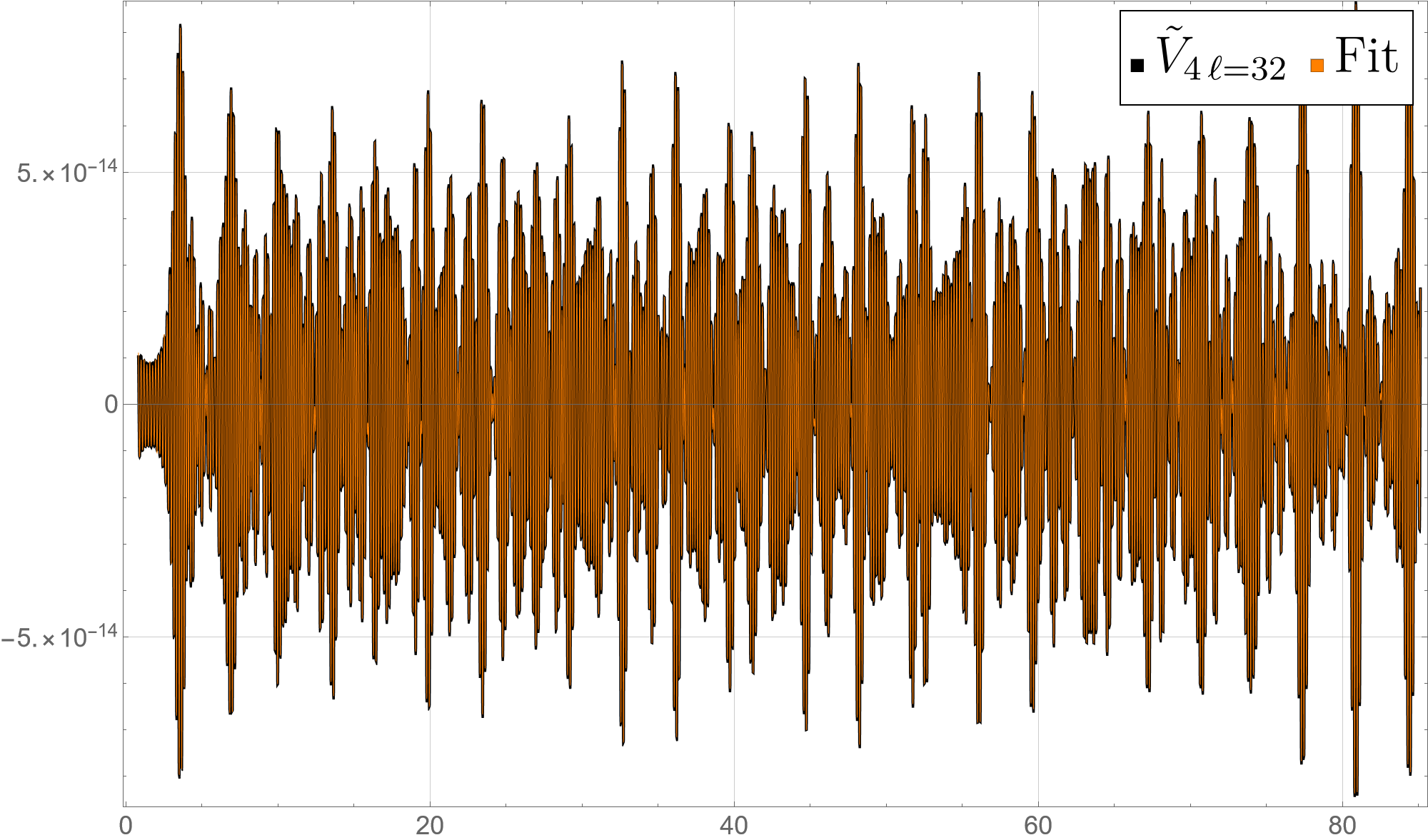} & \includegraphics[width=0.45\linewidth]{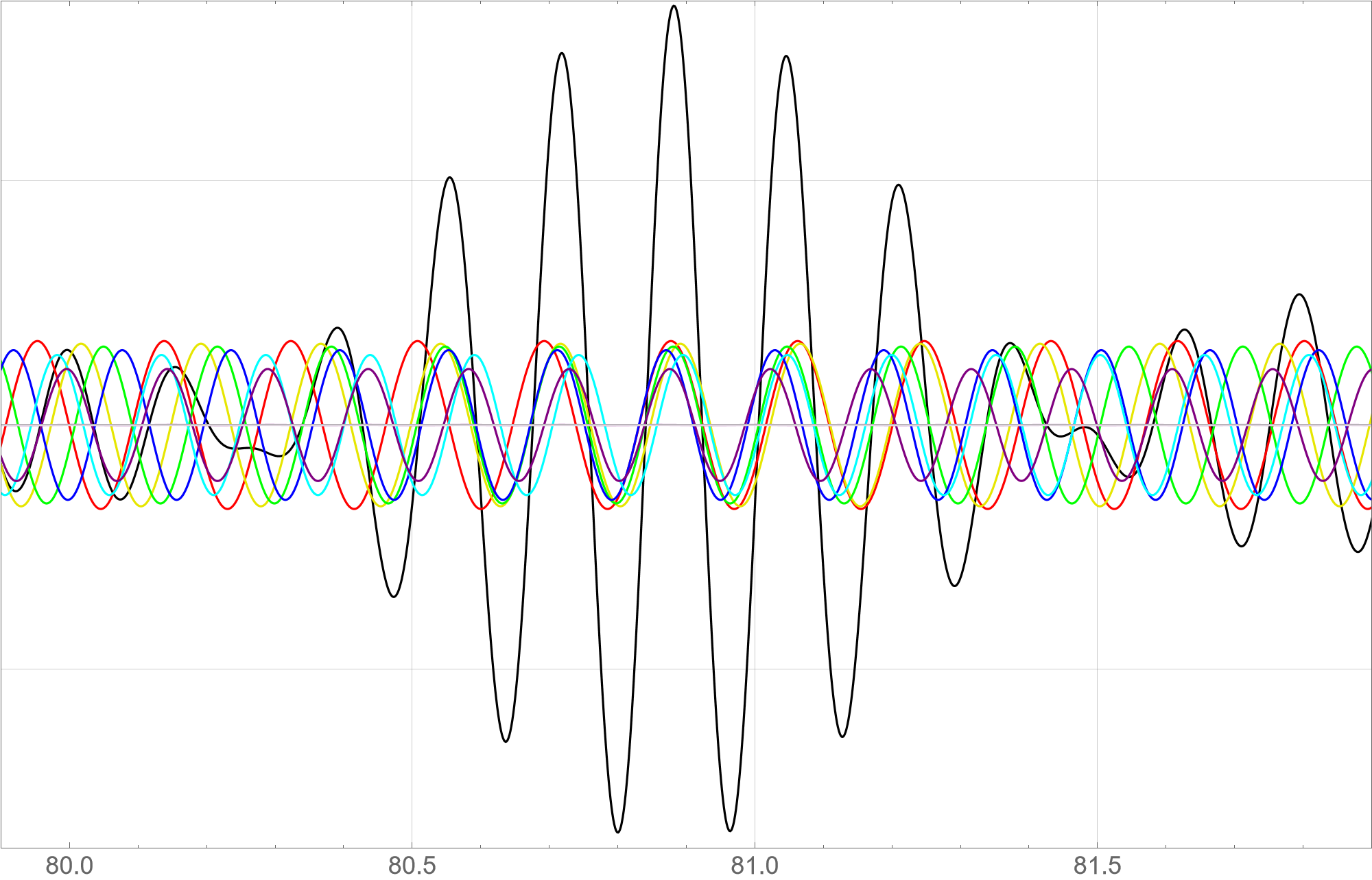}  \\
\multicolumn{2}{c}{\includegraphics[width=0.97\linewidth]{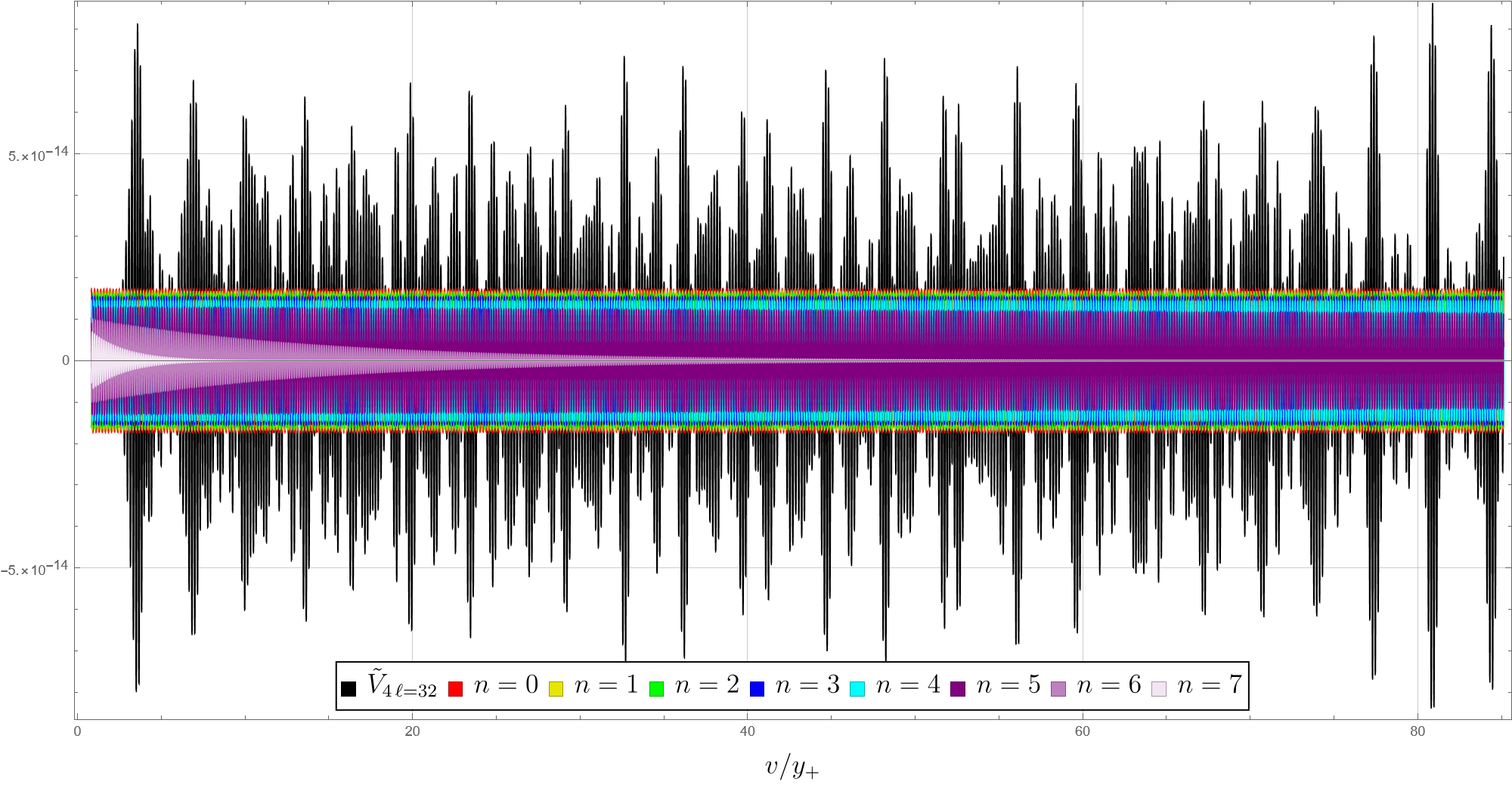}} \\
\multicolumn{2}{c}{\includegraphics[width=0.6\linewidth]{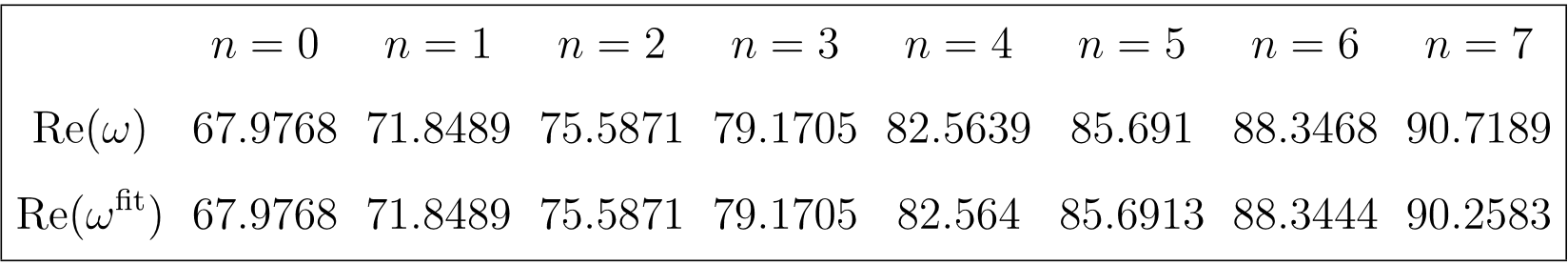}}
\end{tabular}
    \caption{\textit{(Top left)} Plot of the early time evolution of the $\ell=32$ mode of $V_4$ and a fit to a sum of eight QNMs which we identity as the overtones up to $n=7$. \textit{(Bottom)} Plot with overlays showing the overtones. Interference between overtones causes low-frequency beating and large spikes. The real parts of the fitted frequencies $\mathrm{Re}(\omega^{\mathrm{fit}})$ are shown alongside $\mathrm{Re}(\omega)$ from the linearised equations for comparison. \textit{(Top right)} Close-up of a spike produced by overtone interference with overlays showing the overtones.}
    \label{fig:l=32 QNM plots}
\end{figure}

A nontrivial check of the numerical scheme is the observation of QNM behaviour with frequencies and decay rates in agreement with the solutions of the linearised Einstein equations. Fig. \ref{fig:QNM plots} illustrates this agreement for $\ell=2,3,4,5$. We also observe higher overtones with frequencies and decay rates in agreement with the linearised equations. This is shown in Fig. \ref{fig:l=11 QNM plots} for $\ell=11$ where we can resolve the  $n=0,1,2$ overtones in the early time evolution and in Fig. \ref{fig:l=32 QNM plots} for $\ell=32$ where we can resolve up to the $n=7$ overtone. It has been conjectured that resonances in the QNM spectrum may lead to nonlinear instabilities \cite{Moschidis_2023}, so accurately capturing overtones in the evolution is an important validity check of the assumption that higher overtones do not contribute to the predicted late time behaviour \eqref{eqn:prediction} within the time frames explored.

\begin{figure}[p]
    \centering
    \includegraphics[width=1.0\linewidth]{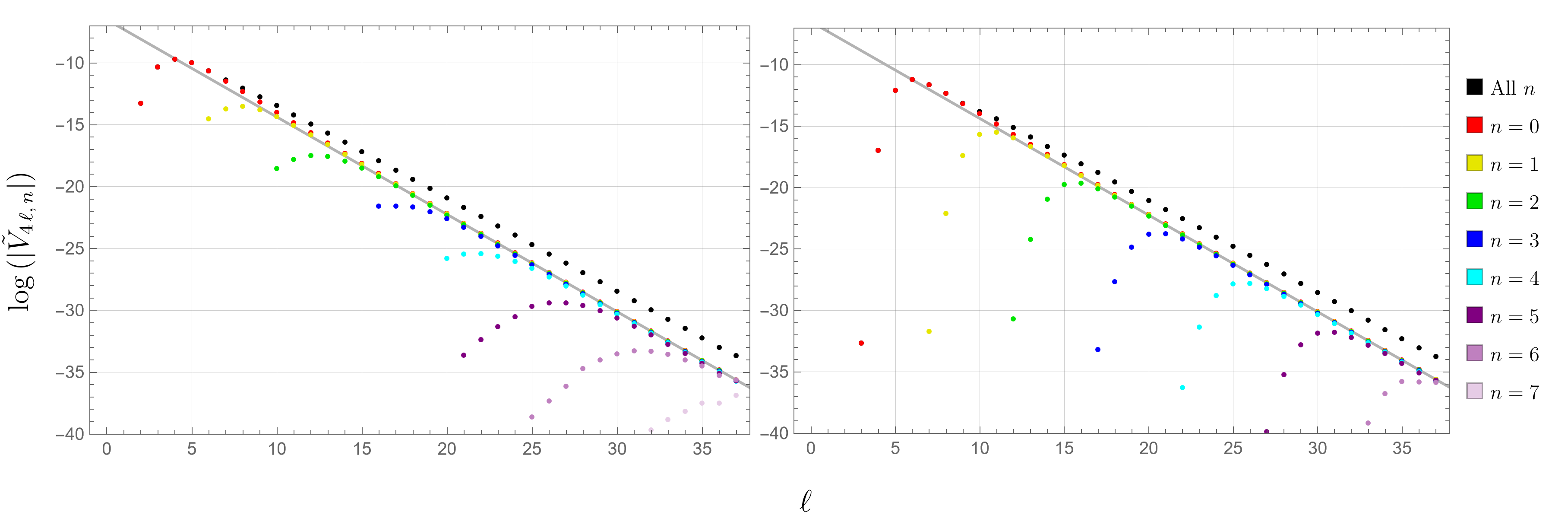}
    \caption{\textit{(Left)} Plot of the spectrum of each resolvable overtone of $V_4$ at time $v=8$. The amplitude of $\tilde{V}_{4\, \ell}$ is shown in black. A straight line fit to the tail of the $n=0$ spectrum with fitted gradient $-\alpha=-0.787$ is shown in grey. \textit{(Right)} Plot at time $v=80$. The peak of the $V_4$ spectrum is dominated by the $n=0$ overtone at both times.}
    \label{fig:overtone spectra}
\end{figure}

The predicted power law decay $|| V_4 ||^2(v) \propto v^{-\frac{2\alpha}{C}}$ depends on the slope $-\alpha$ of the tail of the $V_4$ spectrum. This assumes that the tail is dominated by the fundamental $n=0$ modes. However, Fig. \ref{fig:overtone spectra} shows the $V_4$ spectrum and overtone spectra at early times and we see that this assumption does not hold. We see more and more overtones contributing to the spectrum as $\ell$ increases, and the overtone contributions are not small compared to the $n=0$ mode. Like the $n=0$ modes, the lifetimes of higher $n$ modes increase exponentially with $\ell$ so we expect them to still be prominent in the tail at late times. Despite this, we still expect the prediction to hold because the peak of the spectrum is dominated by the $n=0$ mode and it is the modes around the peak of the spectrum that most contribute to $||V_4||^2$. As the peak of the $n=0$ spectrum moves to higher $\ell$, the peaks of the higher $n$ spectra also move to higher $\ell$ and the peak of the $V_4$ spectrum remains dominated by the fundamental mode. A consequence of this is that the slope of the spectrum $-\alpha$ in the power law prediction should be taken from the tail of the $n=0$ spectrum rather than the full $V_4$ spectrum.

\begin{figure}[p]
\begin{tabular}{cc}
\multicolumn{2}{c}{\includegraphics[width=0.94\linewidth]{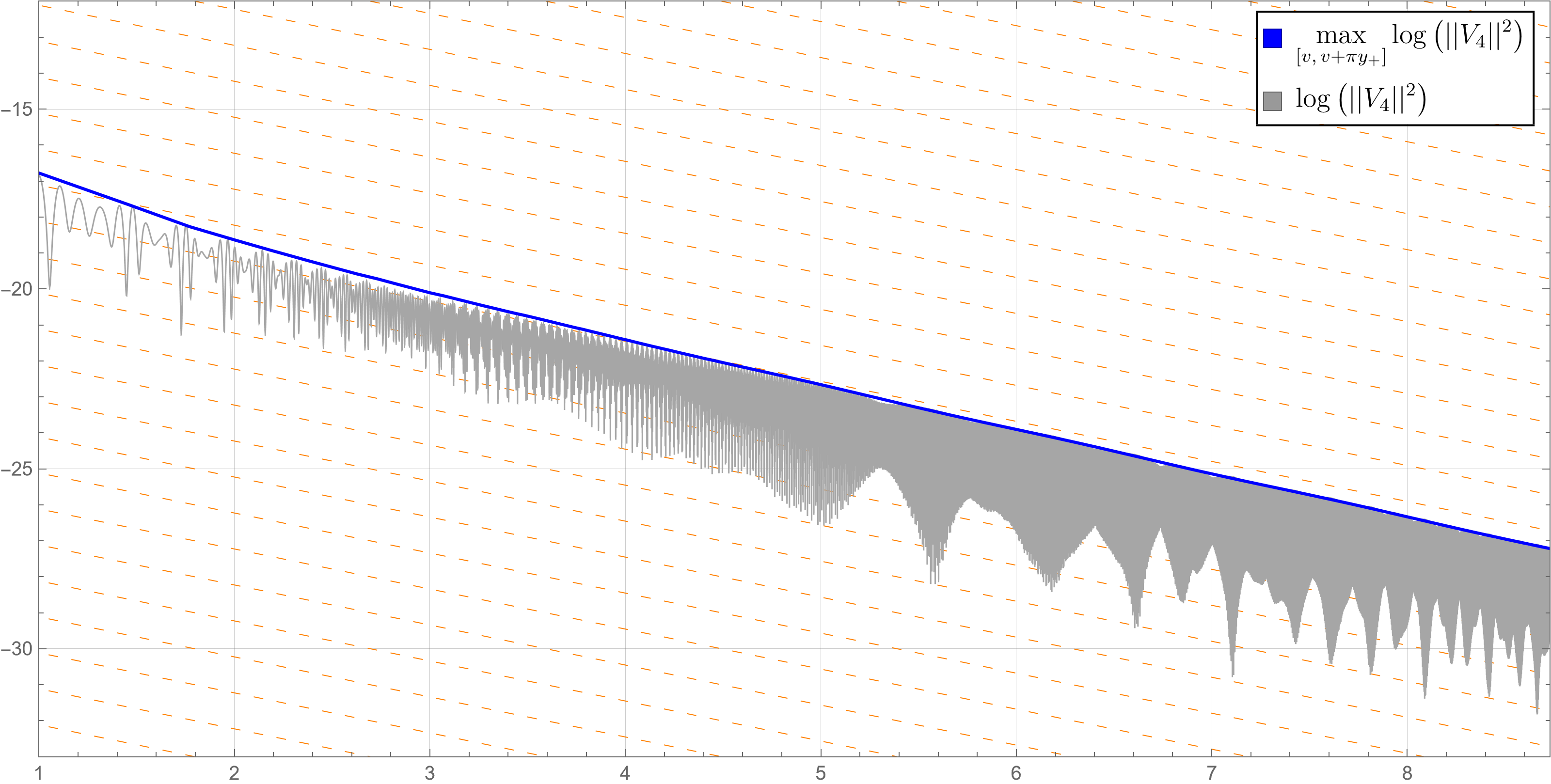}} \\
\multicolumn{2}{c}{\includegraphics[width=0.97\linewidth]{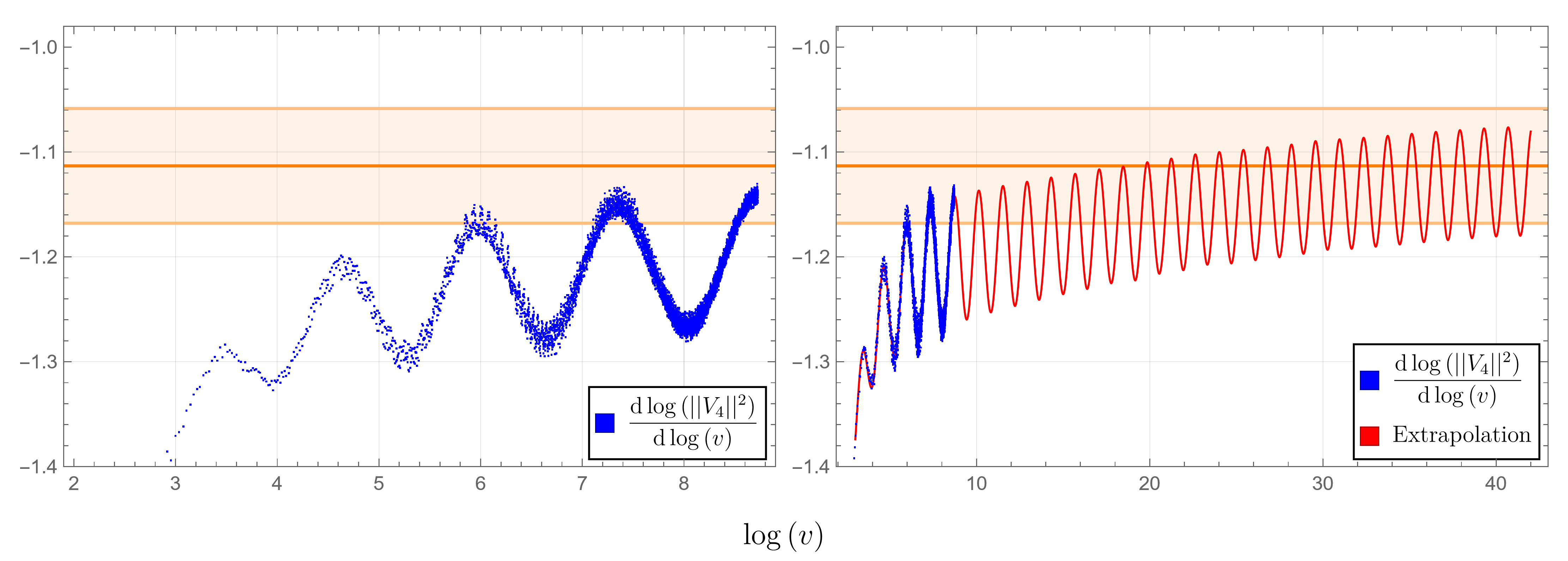}}
\end{tabular}
    \caption{\textit{(Top)} Plot of $\log{(||V_4||^2)}$ against $\log(v)$ for the first $4000$ AdS crossing times for $y_{+}=0.5$. The blue curve shows the maximum over each crossing time. Orange dashed straight lines with gradient equal to the predicted power law exponent are shown for comparison. \textit{(Bottom left)} Plot of the derivative of the maximum of $\log{(||V_4||^2)}$ over each crossing time with respect to $\log(v)$. The shaded orange region shows the mean and amplitude of the predicted oscillatory behaviour. \textit{(Bottom right)} Plot showing an extrapolation to the next $10^{18}$ crossing times using the early time QNM decomposition of $V_4$.}
    \label{fig:full evolution power}
\end{figure}

A straight line fit to the tail of the $n=0$ spectrum on a log plot for this choice of initial data is shown in Fig. \ref{fig:overtone spectra}. The fitted slope is $-\alpha=-0.787$. The predicted late time power law exponent is then approximately $-\frac{2\alpha}{C} = -1.11$. The subleading oscillatory term in \eqref{eqn:loglog grad pred}, which is periodic in $\log(v)$ rather than $v$, has a predicted amplitude of $0.0547$ and period of $C(y_+)=1.41$. Fig. \ref{fig:full evolution power} shows the evolution of $\log(||V_4||^2)$ against $\log(v)$ up to $\log(v)=8.75$, which is over $4000$ AdS boundary crossing times (a single boundary crossing time in $v$ is given by $\pi y_{+}$). $||V_4||^2$ becomes highly oscillatory with time as the solution becomes dominated by modes of higher $\ell$ which oscillate with higher frequency. The prediction for the decay ignores the oscillation of the modes and uses instead the mode amplitudes, so we also plot the maximum of $\log(||V_4||^2(v))$ over each crossing time interval $[v,\, v+\pi y_{+}]$.

We observe that the plot does appear to approach approximately straight line behaviour, indicative of power law decay. Fig. \ref{fig:full evolution power} also shows a plot of the derivative with respect to $\log(v)$. The derivative appears to approach the predicted value of $-\frac{2\alpha}{C}=-1.11$. We also observe oscillations approaching the predicted oscillatory behaviour of this derivative \eqref{eqn:loglog grad pred} with an amplitude and period close to the predicted values.

Although the results of this numerical solution suggest an approach to the predicted behaviour, we note that they do not fully realise the prediction within the timeframe explored. In particular, the gradient of the log-log plot shown in Fig. \ref{fig:full evolution power}, whilst showing a period and amplitude close to the prediction, has not yet reached the predicted mean value, representing the predicted power law exponent, by the end of the timeframe. To explore this further, we extrapolate the results up to $\log{(v)}=42$ using the QNM decomposition of $V_4$ from the early time solution (the first 30 crossing times), including the resolvable overtones. Fig. \ref{fig:full evolution power} shows this extrapolation as well as the result from the nonlinear evolution over the full timeframe. The extrapolation being in good agreement with the nonlinear result suggests that nonlinear effects do not play a dominant role in the late time evolution, at least for the timescales and resolutions explored here. Should the QNM behaviour persist, we see that $||V_4||^2$ would continue to approach the predicted power law behaviour with the predicted subleading oscillations, albeit slowly. This extrapolation shows that it is unsurprising that the numerical results do not fully realise the prediction but only approach it, as the timescale required (the extrapolation shows the next $10^{18}$ crossing times) is beyond the $4000$ crossing times of our numerical evolution.

\begin{figure}[h]
    \includegraphics[width=1.0\linewidth]{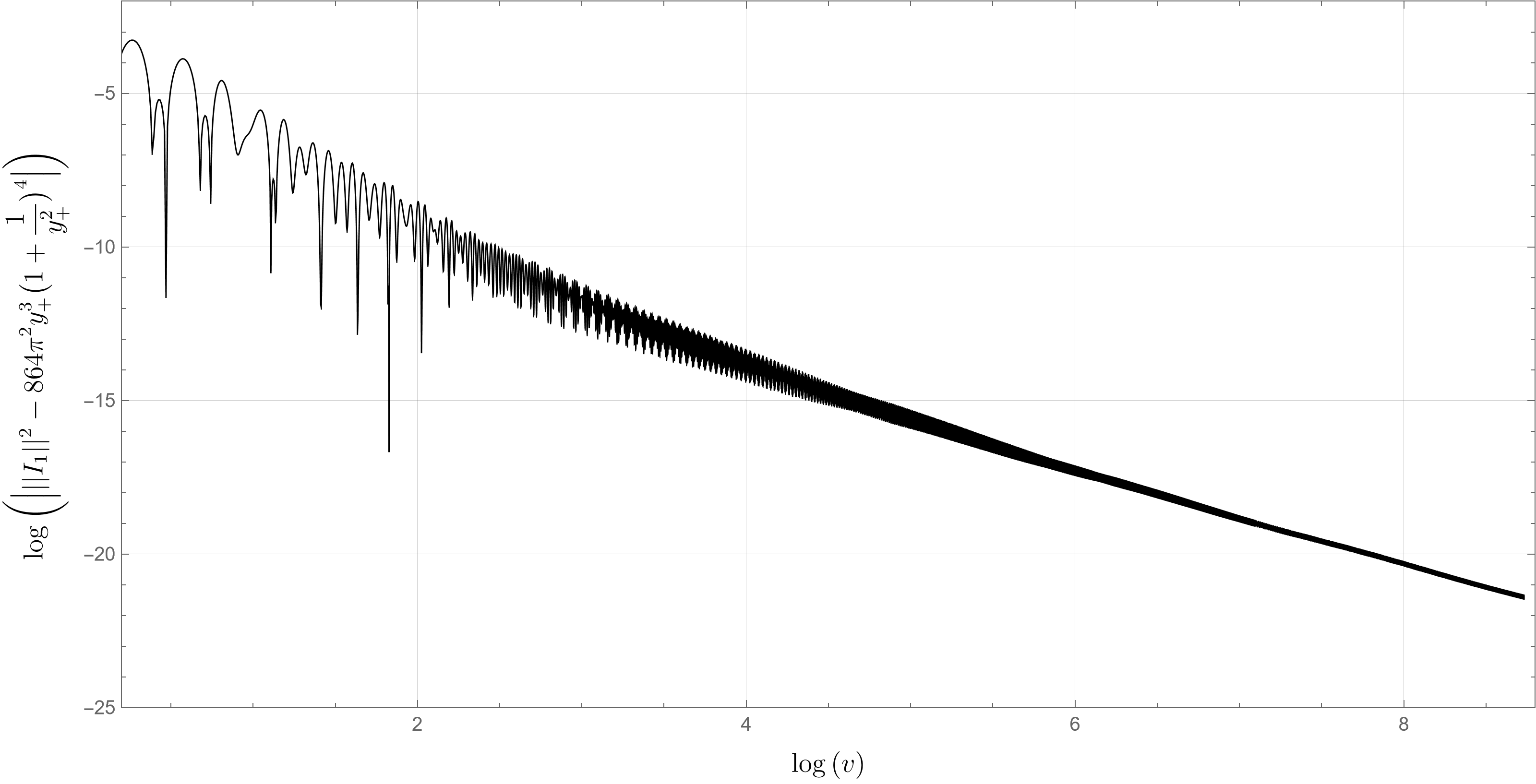}
    \caption{Plot of the decay of $||I_1||^2$ for the first 4000 AdS crossing times for $y_{+}=0.5$, with the exact Schwarzschild-AdS value subtracted. The approach to a straight line at late times suggests a regime of power law decay.}
    \label{fig:y=0.5 Weyl}
\end{figure}

The decay of $||I_1||^2$, defined in \eqref{eqn:I1 def}, is shown in Fig. \ref{fig:y=0.5 Weyl}. The norm $||I_1||^2$ is the quadratic scalar of the Weyl tensor given by $I_1=C_{abcd}C^{abcd}$, squared and integrated against the volume form from the AdS boundary to the apparent horizon on each slice, capturing a more global notion of decay than the boundary quantity $||V_4||^2$. The approach at late times of the log-log plot to a straight line suggests that $||I_1||^2$ also approaches power law behaviour.

The notable slowness of the approach to the prediction is to be expected. It is not due to the presence of long lived $n\geq 1$ overtones at high $\ell$ as we find that the extrapolation shown in Fig. \ref{fig:full evolution power}, which includes the overtones, is indistinguishable from an extrapolation with only the $n=0$ modes. Instead, as described in subsection \ref{sec:how late}, the slowness is due to the difference between the QNM decay rates and the WKB result describing the decay rates at large $\ell$. The predicted power law decay assumes that the QNM decay rates degenerate exponentially with $\ell$, going as $-\operatorname{Im}(\omega_\ell) = \exp{(-C \ell+\kappa)}$. Whilst this is true asymptotically as $\ell\to \infty$, the approach to this asymptotic behaviour with $\ell$ is slow as illustrated in Fig. \ref{fig:decay rate ratios}. Since the mode number $\ell$ at the peak of the $V_4$ spectrum moves with $C(y_{+})^{-1}\log{(v)}$, we expect to wait for an exponentially long time to see convergence to the predicted behaviour.

For $y_{+}=0.5$ we used $N_x$=42 collocation points in the angular direction and $N_z=77$ points in the radial direction with element boundaries at $z=\{ 0, 0.1,0.2,0.3,0.4,0.6,0.8,1\}$ with $11$ points per element for the long time evolution. The early time higher resolution evolution used $N_x=70$ and $N_z=144$ with element boundaries at $z=\{0,0.05,0.1,0.15,0.2,0.25,0.3,0.35,0.4,0.5,0.6,0.8,1\}$ with $12$ points per element.

\newpage

\subsection{\texorpdfstring{A large black hole: $y_{+}=1.0$}{A small black hole: y+ = 1.0}}

We choose initial data $Q_1(0,z,x)=0.5 (1-x^2)(z-z^2)(1+3(x-1)^8+(z-\frac{1}{2})^2)^{-1}$, $V_4(0,x)=0$, $U_4(0,x)=0$ for a black hole with $y_{+}=1.0$. We fix $\chi_4(0,x)$ such that $z=1$ is an apparent horizon at $v=0$. Aside from requiring analyticity in the coordinate patch, the choice of initial data is essentially arbitrary. We make this choice for $y_{+}=1.0$ so that modes up to $\ell=60$ are large enough to be resolvable within machine precision. This perturbation is larger than that used for $y_{+}=0.5$ to allow nonlinear interactions to have the potential to have a greater effect.

\begin{figure}[h]
    \centering
    \includegraphics[width=0.9\linewidth]{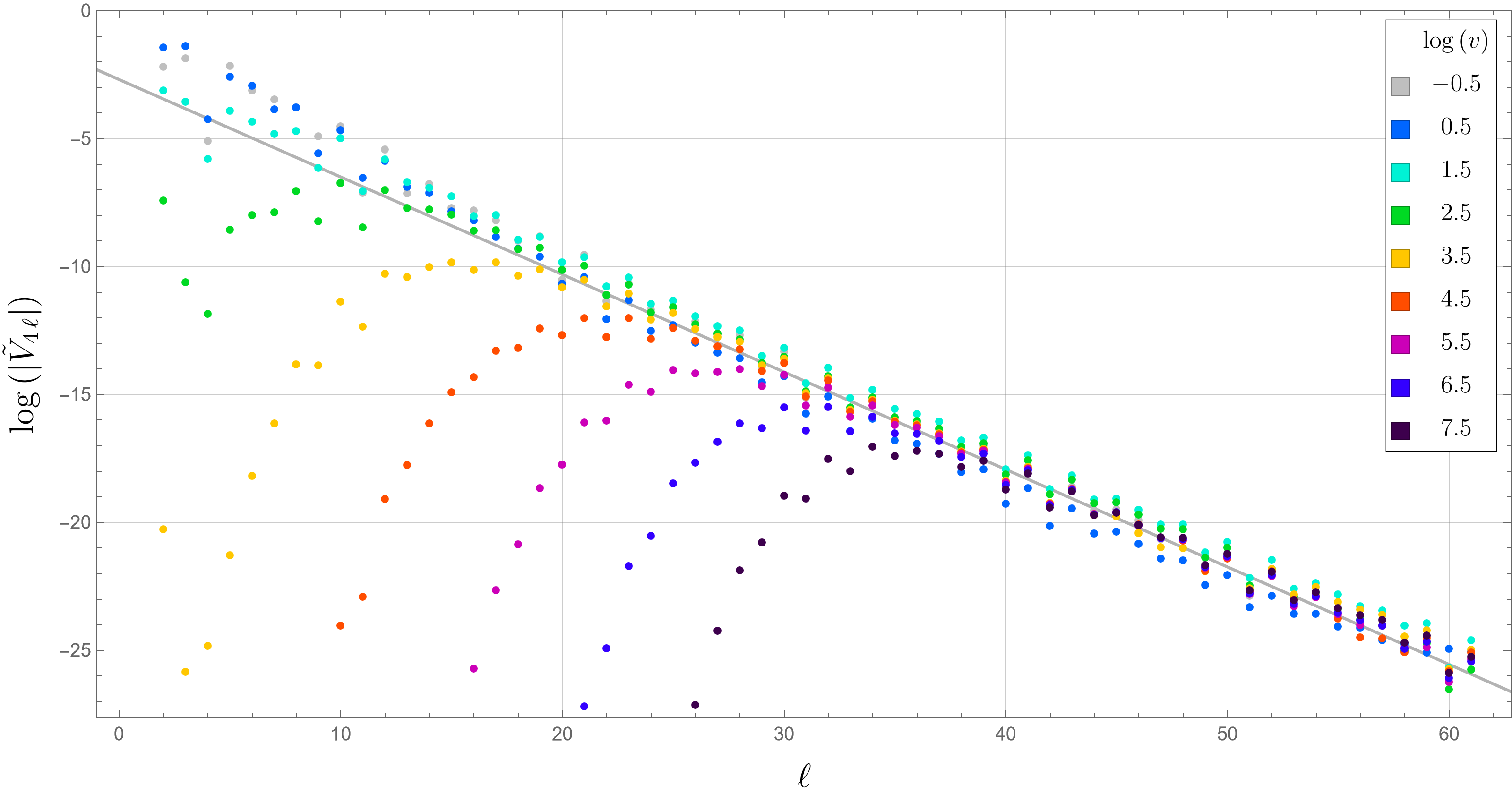}
    \caption{Snapshots of the $V_4$ spectrum for $y_{+}=1.0$ at equally spaced times in $\log(v)$. The peak of the spectrum moves to higher $\ell$ linearly with $C(y_{+})^{-1}\log(v)$ at late times. A straight line fit to the tail of the spectrum, shown in grey, gives a gradient of approximately $-\alpha=-0.38\pm 0.02$. This is used to predict the power law exponent, which is $-\frac{2\alpha}{C}=-2.47\pm0.13$. Each $|\tilde{V}_{4 \, \ell}|$ shown is the maximum of $|\tilde{V}_{4 \, \ell}|$ over a single oscillation of that mode around the time shown.}
    \label{fig:y=1 spectrum}
\end{figure}

Snapshots of the $V_4$ spectrum at equally spaced times in $\log{(v)}$ are shown in Fig. \ref{fig:y=1 spectrum}. We observe the $\ell$ at the peak of the spectrum increasing approximately linearly with $\log{(v)}$ as expected. Unlike the $y_{+}=0.5$ case, we find that for $y_{+}=1.0$ the overtone contributions to the spectrum are small compared to the fundamental $n=0$ modes. We extract the slope of the tail of the spectrum with a linear fit, which is approximately $-\alpha=-0.38\pm 0.02$. Together with $C(y_{+})=0.307$, this leads to a prediction for the late time power law exponent of $-\frac{2\alpha}{C}=-2.47\pm0.13$. The prediction for the gradient of a log-log plot \eqref{eqn:loglog grad pred} also features a subleading oscillatory term, however the predicted oscillation amplitude in this case is $3.4\times10^{-10}$. This is negligible compared to the constant term so we do not expect to observe oscillations. 

\begin{figure}[p]
    \centering
    \includegraphics[width=1.0\linewidth]{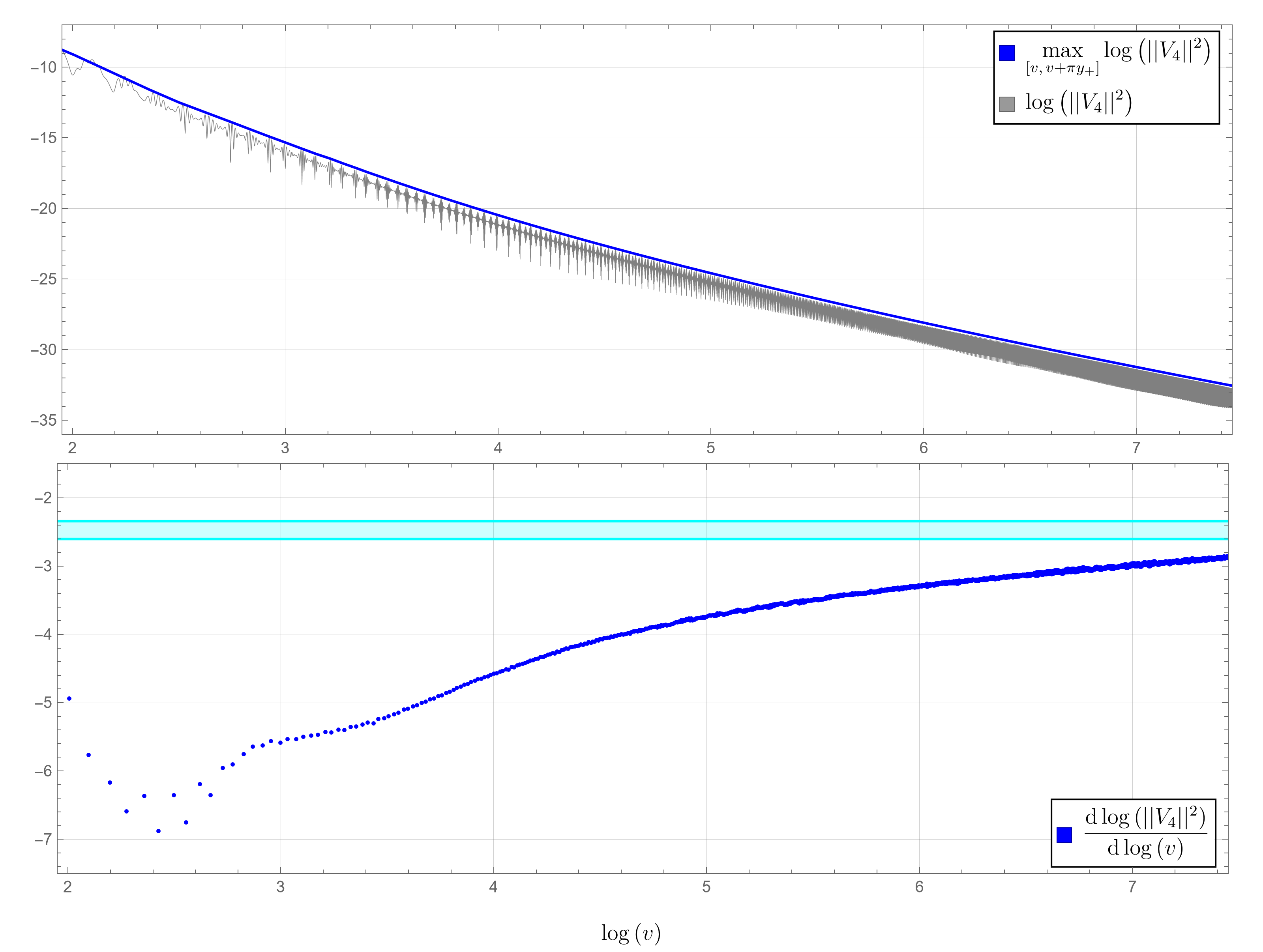}
    \caption{\textit{(Top)} Plot of $\log{(||V_4||^2)}$ against $\log{(v)}$ for $y_{+}=1.0$ for the first 550 AdS crossing times. The maximum over each crossing time is shown in blue. \textit{(Bottom)} The derivative of the maximum of $\log{(||V_4||^2)}$ over each crossing time with respect to $\log{(v)}$ is shown in dark blue. The prediction window for the power law exponent, $-\frac{2\alpha}{C}=-2.47\pm0.13$, is shown in light blue. The predicted amplitude for subleading oscillations is negligible and they are not observed.}
    \label{fig:y=1 power}
\end{figure}

A plot of $\log{(||V_4||^2)}$ against $\log{(v)}$ is shown in Fig. \ref{fig:y=1 power} for the first 550 AdS crossing times. Alongside is a plot of the slope of the log-log plot, where the slope is calculated with respect to the maximum of $\log{(||V_4||^2)}$ over each AdS crossing time. This plot also shows the prediction window for the power law exponent $-\frac{2\alpha}{C}=-2.47\pm0.13$. This plot shows the slope approaching the prediction window with time, which suggests an approach to the predicted power law behaviour. We do not observe any oscillatory behaviour like that seen in the $y_{+}=0.5$ case, but this is expected as the predicted oscillation amplitude is negligible. 

The $V_4$ spectrum at low $\ell\lesssim 20$ does not follow the same exponential slope as the tail of the spectrum, which contributes to a deviation from the predicted behaviour at early times. As in the $y_{+}=0.5$ case, the slowness of the approach to the predicted power law at late times is expected and stems from the slow approach of the QNM decay rates to the asymptotic behaviour, as illustrated in Fig. \ref{fig:decay rate ratios}.

\begin{figure}[p]
    \includegraphics[width=1.0\linewidth]{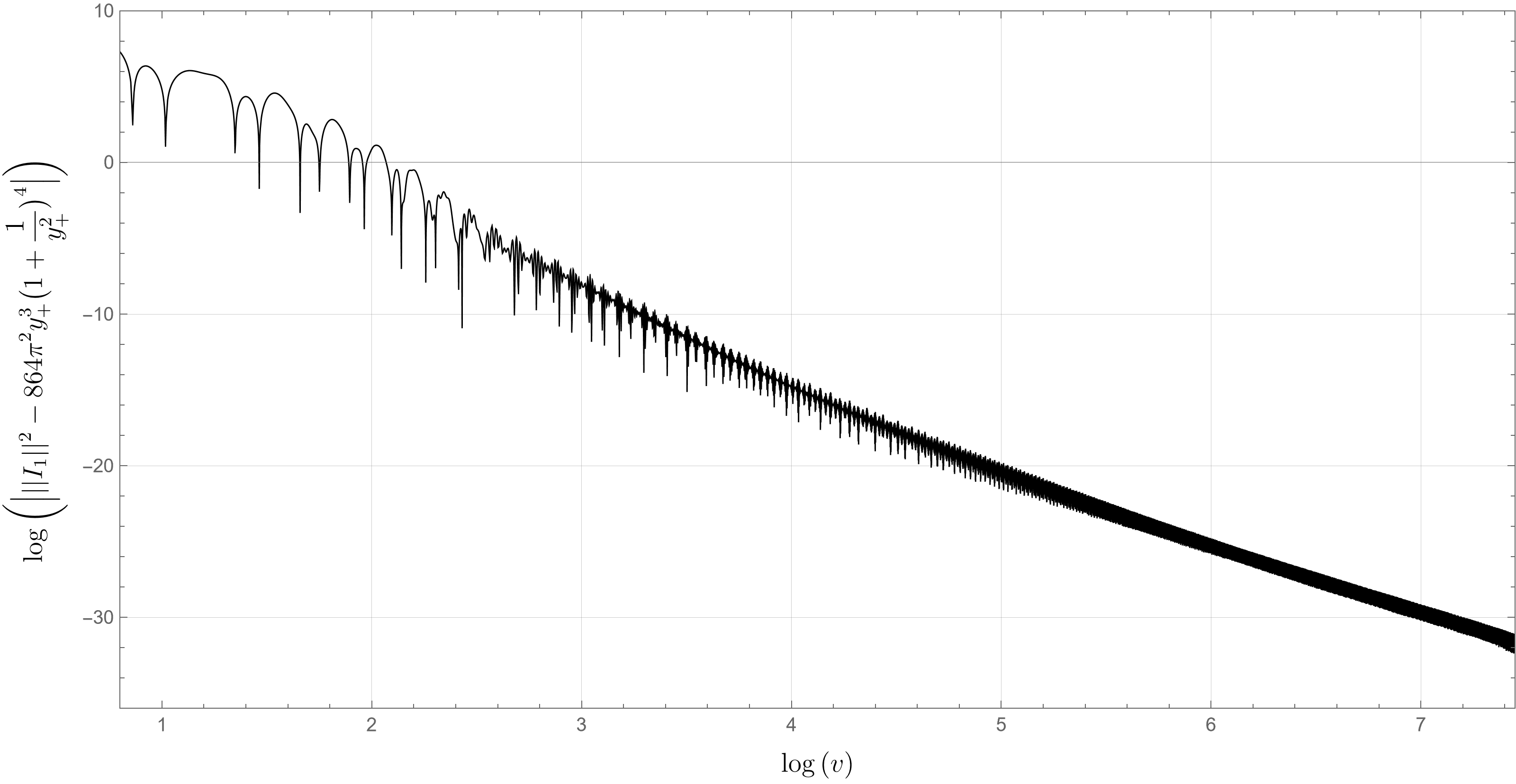}
    \caption{Plot of the log of $||I_1||^2$ with the exact Schwarzschild-AdS value subtracted against $\log{(v)}$ for the first 550 crossing times for $y_{+}=1.0$.}
    \label{fig:y=1 Weyl}
\end{figure}

A log-log plot of $||I_1||^2$ is shown in Fig. \ref{fig:y=1 Weyl}. At early times, $||I_1||^2$ is much larger than for the $y_{+}=0.5$ case due to the larger initial perturbation. At late times, we see an approach to straight line behaviour which indicates an approach to a power law.

For $y_{+}=1.0$ we used $N_x=100$ collocation points in the $x$-direction. The $z$-direction grid used $N_z=84$ points with element boundaries at $z=\{ 0,0.1, 0.2, 0.3, 0.4, 0.6, 0.8, 1 \}$.

\section{Discussion} \label{sec:discussion}
Our simulations provide the first fully nonlinear evidence that, for analytic perturbations of five-dimensional asymptotically AdS black holes that respect an $SO(3)$ isometry, the late-time dynamics enter a clean power-law regime. Within the evolution times and spectral resolutions we could afford, no sign of exponential tails or turbulent cascades appeared; instead, the dominant multipoles settled onto power-law decay exponents that agree with linear predictions when projected onto the same regularity class. Taken together with recent linear-mode analyses, our results strengthen the case that a sizeable sector of small, smooth perturbations is non-linearly stable in $d=5$.

At the same time, the study highlights the limits of current numerical reach. Because the runs terminate before genuine asymptotic times and cannot resolve arbitrarily high angular harmonics, our data do not exclude a change of regime at still later times or at mode numbers beyond our truncation. Likewise, the focus on a couple of horizon-radius–to–AdS-radius ratio $y_+$ - together with relatively small amplitudes—means that the conclusion should be read as a proof of principle, not a universal statement about all black holes in AdS. We believe that the agreement with our simple estimates, based on the large-$\ell$ behaviour of the QNM spectrum of Schwarzschild-AdS black holes, is a strong indication that our analysis has already entered the late-time regime.

Future work can proceed along several directions. One natural extension is to vary the geometric control parameter $y_+$, exploring both smaller and larger black holes to test whether the observed power-law decay persists. In particular, small $y_+$ combined with larger perturbations may bring the system closer to the conjectured regime of nonlinear instability and weak turbulence. Another important direction is to move beyond analytic initial data by considering profiles of lower regularity, \emph{i.e.} finite Sobolev norm, or compactly supported functions. This would allow for a test of whether decay rates continue to follow the predictions of linear theory based on regularity class, or whether singular features give rise to new dynamical phenomena.

Relaxing the imposed $SO(r)$ symmetry offers a further opportunity to uncover richer dynamics. Allowing for only axial symmetry or even fully generic perturbations would enable the study of instability channels like bar modes or superradiance, which are known in five-dimensional asymptotically flat settings and may be enhanced by the reflective boundary of AdS. Additionally, it would be valuable to explore whether similar late-time behaviour occurs in four-dimensional AdS. The linear spectrum in 4D is, to a certain extent, less resonant, which could leave an imprint on the late-time evolution.

Another compelling avenue is to study the response to external boundary sources. By driving particular modes or resonances from the boundary, one could assess whether the system remains stable under external forcing. Stability in this case would represent a stronger form of robustness, while observed instabilities could suggest new mechanisms not captured by passive evolutions. Finally, adding matter fields—such as scalars or gauge fields—would broaden the physical relevance of the analysis and test whether the nonlinear dynamics of pure gravity generalize to more complex systems, especially in holographic contexts.

Ultimately, embedding the calculation in ten-dimensional supergravity would connect directly to $\mathcal{N}=4$ SYM via the AdS/CFT correspondence. The extra spherical directions introduce Kaluza–Klein towers that could either stabilise or destabilise the background, depending on how energy cascades between the AdS$_{5}$ and $S^5$ sectors.

\section*{Acknowledgements}
We are grateful to Christoph~Kehle for his comments on an earlier draft and for helpful discussions regarding his upcoming work with Georgios~Moschidis on the nonlinear stability of four-dimensional Schwarzschild–AdS under rough initial data \cite{KehleMoschidis}. JRVC was supported by a Science and Technology Facilities Council (STFC) studentship ST/W507350/1 and the Cambridge Trust. The work of JES was partially supported by STFC consolidated grant ST/X000664/1 and by Hughes Hall College.

\appendix

\section{Details of numerical method}
\label{appendix:numerics}

\subsection{Polar direction}

For our numerical method, we discretize the spacetime. In the $x$-direction, we use $N_x$ Gauss-Chebyshev-Lobatto collocation points $x_{k}$, $k=0,1,...\,,N_x-1$, given by

\begin{equation}
    x_{k} = \cos{ \left( \frac{k \, \pi}{N_x -1} \right)}.
\end{equation}

\noindent Discretizing functions on this choice of grid points is equivalent to a spectral decomposition into Chebyshev polynomials of the first kind, $T_n (x)$, up to order $n=N_x-1$. 

It would seem more natural to expand functions into Chebyshev polynomials of the second kind, $U_n (x)$, as these are eigenfunctions of the Laplacian on $S^3$, however we find that the $T_n$ basis exhibits better stability under time evolution.

Derivatives of a function $f(x)$ are approximated with derivatives of the truncated polynomial expansion. Denoting the values of the approximation to $f$ at the grid points $x_k$ as $f(x_k)=f_k$, we can calculate the approximate derivatives of $f$ with matrix multiplication. The first derivative $f'_k$ can be computed as $f'_k=\sum_{j}D_{kj}f_j$ where the differentiation matrix $D_{kj}$ has off-diagonal entries

\begin{equation}
    D_{kj} = -\frac{1}{2} \frac{c_j}{c_k} \csc{\left( \frac{(k+j)\pi}{2(N_x-1)} \right)} \csc{\left( \frac{(k-j)\pi}{2(N_x-1)} \right)}, \quad k \neq j
\end{equation}

\noindent where $c_{k}=(-1)^k(1-\frac{1}{2}(\delta_{k,0}+\delta_{k,N_x-1}))$. The diagonal entries are given by

\begin{equation}
    D_{kk} = - \sum_{j \neq k} D_{kj}.
\end{equation}

\noindent Higher order derivatives may be computed by taking matrix powers of $D_{kj}$.

The only nontrivial step in the iteration in the $x$-direction is solving the second-order elliptic equation at $z=1$ for the horizon value of $V$ on each time slice. In terms of the new variable $q_6$, this equation takes the form

\begin{equation}
    (1-x^2)\partial_x\partial_x q_6 +\left(-3x+F_1 \right)\partial_x q_6 + F_0 q_6 = J \label{eqn:elliptic}
\end{equation}

\noindent where $F_1$, $F_0$, and $J$ are functions of the metric variables at $z=1$ which are known by this step in the iteration. We solve this by replacing the derivatives with differentiation matrices, and the functions with their discretized approximations. This gives us a linear system which we can solve for $q_6$ with standard methods. 

Despite this being a second-order elliptic equation, we do not need to provide additional boundary conditions because we are solving for a polynomial approximation to $q_6$ which forces regularity at $x=\pm 1$. This guarantees that $\partial_x \partial_x q_6$ is finite at $x=\pm 1$, so evaluating \eqref{eqn:elliptic} at $x=\pm 1$ fixes the Robin boundary conditions there.

\subsection{Radial direction}

In the $z$-direction, we employ a spectral element discontinuous Galerkin (DG) method. We subdivide the domain into $N_e$ elements, with the $i^{\mathrm{th}}$ element $\Omega_i$ having $N^{(i)}_z$ Gauss-Legendre-Lobatto collocations points. The total number of collocation points in the $z$-grid is then $N_z=\sum_{i}N^{(i)}_z$. We denote the grid points by $z^{(i)}_{k}$, where $k=1,...\,,N^{(i)}_z$ and $i=1,...\,, N_e$. 

Functions $f(z)$ are approximated by their values at the grid points $f(z^{(i)}_{k})=f^{(i)}_{k}$, which is equivalent to decomposing the function in the $i^{\mathrm{th}}$ element into a basis of scaled Legendre polynomials $P_{n}$, truncated at order $n=N^{(i)}_z-1$. We denote the $m^{\mathrm{th}}$ global basis function by $v_m$, which is a $P_n$ compactly supported on one of the elements. The values of $z$ at the boundaries between elements have grid points in both of the adjoining elements, $z^{(i)}_{N^{(i)}_z}=z^{(i+1)}_{1}$, however in the DG method we do not force $f^{(i)}_{N^{(i)}_z}=f^{(i+1)}_{1}$.

Solving for the metric components $p_i$, $i=1,...\,, 5$, at each timestep $v$ requires solving ODEs of the form

\begin{equation}
    \partial_z p(v,z,x) + \frac{F(v,z,x)}{z} p(v,z,x)=J(v,z,x). \label{eqn:ODE}
\end{equation}

\noindent In the DG method we approximate $p$ with truncated Legendre polynomial expansions in each element that satisfy the weak form of \eqref{eqn:ODE}

\begin{equation}
    -\int_{\Omega_i} v_m' \, p + \left[ v_{m} \, \hat{p} \right]_{\partial \Omega_i}+ \int_{\Omega_i} v_m \, \frac{F}{z} \,p = \int_{\Omega_i}v_m \, J
\end{equation}

\noindent where we have integrated by parts and replaced the boundary term with flux term $\left[ v_{m} \, \hat{p} \right]_{\partial \Omega_i}=v_m(b)\,\hat{p}_b - v_m(a)\,\hat{p}_a$, where $a=z^{(i)}_{1}, b=z^{(i)}_{N_z^{(i)}}$. We choose an upwind numerical flux $\hat{p}_a = p(a_-)$ where $a_-=z^{(i-1)}_{N_z^{(i-1)}}$. The numerical flux in the element that includes the AdS boundary $z=0$ is used to set the Dirichlet boundary conditions \eqref{eqn:Dirichlet BCs first}-\eqref{eqn:Dirichlet BCs last}. Solving for the variables $p_i(v,z,x)=q_i(v,z,x)-q_i(v,0,x)$ instead of $q_i$ sets the $z=0$ flux to zero. Using $p_i$ instead of $q_i$ also leads to better conditioning when dealing with the $p_i \,z^{-1}$ terms.

The final step on each hypersurface is finding $\partial_v Q_1$ which can be reconstructed from the definition of $d_t A$ \eqref{eqn:defdtA}. This takes the form

\begin{equation}
    \partial_v Q_1 = -\partial_zf(Q_1) + J \label{eqn:DvQ1}
\end{equation}

\noindent which allows us to obtain $\partial_v Q_1$ algebraically from known quantities on the hypersurface. However, doing this algebraically is unstable under time evolution. Instead, we solve for $\partial_v Q_1$ using the weak form of \eqref{eqn:DvQ1} given by

\begin{equation}
    \int_{\Omega_i} v_m \,\partial_v Q_1 = \int_{\Omega_i} v_m' \,f(Q_1) - [v_m \hat{f}(Q_1)]_{\partial \Omega_i}+ \int_{\Omega_i} v_m \,J
\end{equation}

\noindent after integrating by parts. The flux term $[v_m \hat{f}(Q_1)]_{\partial \Omega_i}$ must be treated more carefully than the upwind flux used previously because there may be waves propagating in either direction at each element boundary. We employ the Lax-Friedrichs flux, given by

\begin{equation}
    \hat{f}(Q_1) = \frac{1}{2} \left( f(Q_{1-}) + f(Q_{1+})) \right) - \frac{\alpha}{2} (Q_{1+}-Q_{1-}) \label{eqn:LFflux}
\end{equation}

\noindent where $Q_{1-}$ and $Q_{1+}$ denote the numerical values of $Q_1$ at an element boundary. For the flux at an element boundary $z=z^{(i)}_{N_z^{(i)}}$, $Q_{1-}$ denotes $Q_1$ at $z^{(i)}_{N_z^{(i)}}$ and $Q_{1+}$ denotes $Q_1$ in the adjacent element at $z^{(i+1)}_{1}$. The $\alpha$ in \eqref{eqn:LFflux} is given by \begin{equation}
    \alpha = \max_{u\in \{ Q_{1-}, Q_{1+}\} } | f'(u) |.
\end{equation}

\noindent The flux at the AdS boundary $z=0$ is determined by the boundary expansion. We use an upwind flux at the apparent horizon $z=1$, which is consistent with the causal structure of the horizon.

The RK4 timestep $\Delta v$ must be sufficiently small to avoid instabilities. This scales with the size of the smallest grid spacing, given in terms of the radial grid parameters as

\begin{equation}
    \Delta v = \frac{C}{N_{e} \big( \max_{i} N_{z}^{(i)} \big)^2 }
\end{equation}

\noindent where $C$ is the CFL factor. The largest possible $C$ for a given choice of grid can be found empirically. This allows for typically larger timesteps than a spectral method with the same number of collocation points $N_z$ where the timestep scales as $\Delta v \propto N_z^{-2}$. Typically we found CFL factors $C \lesssim2$ to be stable, with long time evolutions using $C=1$ and early time higher resolution evolutions using $C=0.5$. The threshold $\Delta v$ for stability also scales with $N_x$ as $N_x^{-2}$ and linearly with $y_{+}$ however it is primarily the scaling with $N_e$ and $N_z^{(i)}$ that dictates $\Delta v$ for the grids used here.

The DG method has the advantage that we are free to choose where to place the boundaries of each element and how many points per element. This allows us to have a higher resolution near the AdS boundary by placing more points there, without having to have a high resolution everywhere. This is important at late times when the high angular momentum modes are localised near the boundary. Accurate late time evolutions would be computationally expensive and time consuming without this freedom. Another advantage of using finite elements is that the differential operator matrices are sparse and (almost) block diagonal. These are computationally cheaper to store and invert, and can be parallelised on multi-core processors.

\subsection{Aliasing}

The numerical time evolution produces spurious modes known as the aliasing phenomenon. Nonlinear terms in the equations generate modes with mode numbers higher than the highest mode in our discretization, and these are aliased down into the lower modes. This not only produces errors but can also lead to instabilities that destroy the time evolution. To remedy this, we employ the Orszag two-thirds rule in the polar $x$-direction \cite{Boyd_1989}.

We find that implementing the 2/3 rule by discarding the top third of $T_n$ modes of each variable at each timestep leads to a significant buildup of errors around the cutoff mode number. This is because Chebyshev polynomials of the first kind $T_n$ are not the natural basis functions for the metric components. The variables $V_4$ and $\chi_4$ are $S^3$ scalars so their natural basis functions are eigenfunctions of the $S^3$ Laplacian, which are Gegenbauer polynomials $C^{(\alpha)}_{n}(x)$ with $\alpha=1$ (which are Chebyshev polynomials of the second kind). Since $C^{(1)}_n(x)=\sum_{m=0}^{n}T_{|n-2m|}(x)$, if there is a spurious $C^{(1)}_{n}$ mode then deleting the $T_n$ mode only shifts the spurious mode to $C^{(1)}_{n-2}$. Similarly, $U_4$ is naturally expanded in $C^{(2)}_{n}$ and $Q_1$ in $C^{(3)}_{n}$. In our implementation, we dealias by expanding each variable in its natural basis and then applying the 2/3 rule. This is implemented numerically by matrix multiplication. The dealiasing matrices $A^{(\alpha)}_{kj}$ are given by

\begin{equation}
    A^{(\alpha)}_{kj} = \sum_{n=0}^{\lambda} \frac{n! (n+\alpha)\Gamma(\alpha)^2}{2^{1-2\alpha} \Gamma(n+2\alpha)}  w_j \,(1-x_j^2)^{\alpha} \,C^{(\alpha)}_{n}(x_k) C^{(\alpha)}_{n}(x_j)
\end{equation}

\noindent where $w_{j}=\frac{1}{N_x-1}(1-\frac{1}{2}(\delta_{j,0}+\delta_{j,N_x-1}))$, and $\lambda$ is the highest surviving mode after dealiasing which we choose to be $\lambda = \lfloor \frac{2}{3} (N_x-1) \rfloor$. The dealiasing matrices are the Chebyshev-Gauss quadrature approximations to integrals that project out the $C_n^{(\alpha)}$ mode coefficients. Since the integrands are polynomials, for which the quadrature sums exactly equal the integrals, the dealiasing matrices leave modes below the cutoff unchanged.

In the radial $z$-direction, variables are expanded into a basis of compactly supported Legendre polynomials scaled to each element. We dealias at each timestep by discarding only the top mode in each element.

\section{Initial data with an apparent horizon} \label{appendix:AH}

Initial data is specified by choosing $Q_1(v,z,x)$, $U_4(v,x)$, $V_4(v,x)$ and $\chi_4(v,x)$ at $v=0$. We require that there is an apparent horizon at $z=1$, which amounts to the horizon condition \eqref{eqn:horizoncondition} being satisfied. We do this by choosing only $Q_1$, $U_4$ and $V_4$, and then finding $\chi_4$ such that the horizon condition is satisfied. This is done numerically using Newton's method by trying an initial guess for $\chi_4$, calculating the expansion of outgoing null rays at $z=1$, and using the result to modify $\chi_4$. The updated $\chi_4$ is then used as the new guess and the process is repeated until a $\chi_4$ is obtained such that the expansion at $z=1$ is sufficiently close to zero (to within machine precision).

The horizon condition can be rewritten in terms of $d_t \chi$ instead of $\partial_v \chi$ and $V$ so that we only need to carry out the marching orders up to $d_t\chi$ to calculate the expansion. This is given by

\begin{equation}
    \Theta \equiv  \left[ d_t \chi - \frac{1}{2}\left(1+\frac{z^2}{y_{+}^2}\right)\left(\partial_z \chi - \frac{1}{z}\right)+(1-x^2) U \partial_x \chi + \frac{1}{3} \left((1-x^2)\partial_x U - 3 x  U \right) \right] \bigg|_{z=1} = 0.
\end{equation}

\noindent Given $Q_1$, $U_4$, $V_4$, and $\chi_4$, the marching orders can be carried out up to $d_t\chi$ and $\Theta$ can be evaluated. Suppressing the dependence on $Q_1$, $U_4$, and $V_4$, let $\Theta_i (\mathbf{u})$ denote the value of $\Theta$ at $x_i$, $i=1,...,N_x$, and $\mathbf{u}=(u_1,...,u_{N_x})$ denote the vector of values of $\chi_4$ at the collocation points, $u_i=\chi_4(0,x_i)$. The discretised form of the horizon condition is then $\Theta_i(\mathbf{u})=0$. Starting with an initial guess $\mathbf{u}^{(0)}$ for $\mathbf{u}$, Newton's method provides $\mathbf{u}^{(n+1)}$ from $\mathbf{u}^{(n)}$ as the solution to the equation

\begin{equation}
    \Theta_i (\mathbf{u}^{(n)}) + \sum_{j} \frac{\partial \Theta_i (\mathbf{u}^{(n)})}{\partial u_j} \left(u_j^{(n+1)}-u_j^{(n)}\right) = 0. \label{eqn:Newton}
\end{equation}

\noindent This could be solved directly by computing and inverting the Jacobian matrix $\mathrm{J}_{ij}(\mathbf{u}^{(n)})=\partial_{u_{j}} \Theta_i (\mathbf{u}^{(n)})$ but computing the Jacobian is computationally costly so instead we use a Jacobian-free Newton-Krylov method \cite{KNOLL2004357}. This utilises the property that, whilst we don't have $\mathbf{J}(\mathbf{u}^{(n)})$, we can approximate $\mathbf{J}(\mathbf{u}^{(n)})\mathbf{v}$ for $\mathbf{v}\in \mathbb{R}^{N_x}$ using

\begin{equation}
    \mathbf{J}(\mathbf{u}^{(n)})\mathbf{v} \approx \frac{1}{2 \epsilon} \left( \mathbf{J}(\mathbf{u}^{(n)}+\epsilon\mathbf{v}) - \mathbf{J}(\mathbf{u}^{(n)} - \epsilon\mathbf{v}) \right) \label{eqn:Jacobian approx}
\end{equation}

\noindent for some $\epsilon$, $0<\epsilon\ll 1$. Repeated application of \eqref{eqn:Jacobian approx} $k$ times approximates $\mathbf{J}(\mathbf{u}^{(n)})^k\,\mathbf{v}$. Equipped with these approximations, we solve \eqref{eqn:Newton} iteratively using the generalised minimal residual method (GMRES) \cite{GMRES}. Starting with a guess $\mathbf{v}^{(0)}$, the $k^{\mathrm{th}}$ iteration finds the vector $\mathbf{v}^{(k)}$ that minimises the Euclidean norm of the residual 

\begin{equation}
    \mathbf{r}^{(k)}=\frac{1}{||\mathbf{\Theta}(\mathbf{u}^{(n)})||_2} \left( \mathbf{\Theta} (\mathbf{u}^{(n)})-\mathbf{J}(\mathbf{u}^{(n)})\mathbf{v}^{(k)} \right).
\end{equation}

\noindent The minimisation is constrained to $\mathbf{v}^{(k)}$ of the form $\mathbf{v}^{(k)}=\mathbf{v}^{(0)}+\mathbf{w}^{(k)}$, $\mathbf{w}^{(k)} \in K_k$, where $K_k$ is the $k^{\mathrm{th}}$ Krylov subspace, defined by

\begin{equation}
    K_k = \mathrm{span} \{ \mathbf{r}^{(0)}, \,\mathbf{J}(\mathbf{u}^{(n)})\mathbf{r}^{(0)}, \,\mathbf{J}(\mathbf{u}^{(n)})^2\mathbf{r}^{(0)} , ..., \,\mathbf{J}(\mathbf{u}^{(n)})^{k-1}\mathbf{r}^{(0)} \}.
\end{equation}

\noindent This is a least squares minimisation problem that can be solved with a QR decomposition. We choose $\mathbf{v}^{(0)}=0$ for the initial guess. We stop the iteration at $k$ such that $||\mathbf{r}^{(k)}||_2$ is below some threshold. The updated candidate for $\mathbf{u}$ is then $\mathbf{u}^{(n+1)}=\mathbf{u}^{(n)} - \mathbf{v}^{(k)}$. This process is iterated until we reach $n$ such that $||\mathbf{\Theta}(\mathbf{u}^{(n)})||_2$ is sufficiently close to zero.

Newton's method requires the initial guess $\mathbf{u}^{(0)}$ to be close to $\mathbf{u}$ in order to converge to $\mathbf{u}$. The choice of initial data $Q_1(0,z,x)=0$, $U_4(0,x)=V_4(0,x)=\chi_4(0,x)=0$ corresponds to the exact solution Schwarzschild-AdS$_{5}$, so for small perturbations we choose our initial iteration seed for $\mathbf{u}$ to be $\mathbf{u}^{(0)}=0$. For large perturbations, where $Q_1$, $U_4$ and $V_4$ are not small, we find $\mathbf{u}$ iteratively. We do this in $M$ steps by finding $\mathbf{u}$ for the sequence of initial data choices $\frac{m}{M}Q_1$, $\frac{m}{M}U_4$ and $\frac{m}{M}V_4$, $m=1,...,M$, using the $\mathbf{u}$ found at each step $m$ as the seed for the next step $m+1$. We use the seed $\mathbf{u}^{(0)}=0$ for the first step $m=1$.

\printbibliography

@article{Regge:1957td,
    author = "Regge, Tullio and Wheeler, John A.",
    title = "{Stability of a Schwarzschild singularity}",
    doi = "10.1103/PhysRev.108.1063",
    journal = "Phys. Rev.",
    volume = "108",
    pages = "1063--1069",
    year = "1957"
}

@article{Vishveshwara:1970cc,
    author = "Vishveshwara, C. V.",
    title = "{Stability of the schwarzschild metric}",
    doi = "10.1103/PhysRevD.1.2870",
    journal = "Phys. Rev. D",
    volume = "1",
    pages = "2870--2879",
    year = "1970"
}

@article{Price:1971fb,
    author = "Price, Richard H.",
    title = "{Nonspherical perturbations of relativistic gravitational collapse. 1. Scalar and gravitational perturbations}",
    doi = "10.1103/PhysRevD.5.2419",
    journal = "Phys. Rev. D",
    volume = "5",
    pages = "2419--2438",
    year = "1972"
}

@article{10.1063/1.524181,
    author = {Wald, Robert M.},
    title = {Note on the stability of the Schwarzschild metric},
    journal = {Journal of Mathematical Physics},
    volume = {20},
    number = {6},
    pages = {1056-1058},
    year = {1979},
    month = {06},
    issn = {0022-2488},
    doi = {10.1063/1.524181},
    url = {https://doi.org/10.1063/1.524181},
    eprint = {https://pubs.aip.org/aip/jmp/article-pdf/20/6/1056/19001857/1056\_1\_online.pdf},
}

@article{Bardeen:1973xb,
    author = "Bardeen, J. M. and Press, W. H.",
    title = "{Radiation fields in the schwarzschild background}",
    doi = "10.1063/1.1666175",
    journal = "J. Math. Phys.",
    volume = "14",
    pages = "7--19",
    year = "1973"
}

@article{Moncrief:1974am,
    author = "Moncrief, V.",
    title = "{Gravitational perturbations of spherically symmetric systems. I. The exterior problem.}",
    doi = "10.1016/0003-4916(74)90173-0",
    journal = "Annals Phys.",
    volume = "88",
    pages = "323--342",
    year = "1974"
}

@article{Zerilli:1970se,
    author = "Zerilli, Frank J.",
    title = "{Effective potential for even parity Regge-Wheeler gravitational perturbation equations}",
    doi = "10.1103/PhysRevLett.24.737",
    journal = "Phys. Rev. Lett.",
    volume = "24",
    pages = "737--738",
    year = "1970"
}

@article{Dafermos:2016uzj,
    author = "Dafermos, Mihalis and Holzegel, Gustav and Rodnianski, Igor",
    title = "{The linear stability of the Schwarzschild solution to gravitational perturbations}",
    eprint = "1601.06467",
    archivePrefix = "arXiv",
    primaryClass = "gr-qc",
    doi = "10.4310/acta.2019.v222.n1.a1",
    journal = "Acta Mat.",
    volume = "222",
    number = "1",
    pages = "1--214",
    year = "2019"
}

@article{Klainerman:2017nrb,
    author = "Klainerman, Sergiu and Szeftel, Jeremie",
    title = "{Global Nonlinear Stability of Schwarzschild Spacetime under Polarized Perturbations}",
    eprint = "1711.07597",
    archivePrefix = "arXiv",
    primaryClass = "gr-qc",
    month = "11",
    year = "2017"
}

@article{Dafermos:2021cbw,
    author = "Dafermos, Mihalis and Holzegel, Gustav and Rodnianski, Igor and Taylor, Martin",
    title = "{The non-linear stability of the Schwarzschild family of black holes}",
    eprint = "2104.08222",
    archivePrefix = "arXiv",
    primaryClass = "gr-qc",
    month = "4",
    year = "2021"
}

@article{Hintz:2016gwb,
    author = "Hintz, Peter and Vasy, Andr\'as",
    title = "{The global non-linear stability of the Kerr-de Sitter family of black holes.}",
    eprint = "1606.04014",
    archivePrefix = "arXiv",
    primaryClass = "math.DG",
    doi = "10.4310/ACTA.2018.v220.n1.a1",
    journal = "Acta Math.",
    volume = "220",
    number = "1",
    pages = "1--206",
    year = "2018"
}

@article{Maldacena:1997re,
    author = "Maldacena, Juan Martin",
    title = "{The Large $N$ limit of superconformal field theories and supergravity}",
    eprint = "hep-th/9711200",
    archivePrefix = "arXiv",
    reportNumber = "HUTP-97-A097, HUTP-98-A097",
    doi = "10.4310/ATMP.1998.v2.n2.a1",
    journal = "Adv. Theor. Math. Phys.",
    volume = "2",
    pages = "231--252",
    year = "1998"
}

@article{Gubser:1998bc,
    author = "Gubser, S. S. and Klebanov, Igor R. and Polyakov, Alexander M.",
    title = "{Gauge theory correlators from noncritical string theory}",
    eprint = "hep-th/9802109",
    archivePrefix = "arXiv",
    reportNumber = "PUPT-1767",
    doi = "10.1016/S0370-2693(98)00377-3",
    journal = "Phys. Lett. B",
    volume = "428",
    pages = "105--114",
    year = "1998"
}

@article{Witten:1998qj,
    author = "Witten, Edward",
    title = "{Anti de Sitter space and holography}",
    eprint = "hep-th/9802150",
    archivePrefix = "arXiv",
    reportNumber = "IASSNS-HEP-98-15",
    doi = "10.4310/ATMP.1998.v2.n2.a2",
    journal = "Adv. Theor. Math. Phys.",
    volume = "2",
    pages = "253--291",
    year = "1998"
}

@article{Aharony:1999ti,
    author = "Aharony, Ofer and Gubser, Steven S. and Maldacena, Juan Martin and Ooguri, Hirosi and Oz, Yaron",
    title = "{Large N field theories, string theory and gravity}",
    eprint = "hep-th/9905111",
    archivePrefix = "arXiv",
    reportNumber = "CERN-TH-99-122, HUTP-99-A027, LBNL-43113, RU-99-18, UCB-PTH-99-16, LBL-43113",
    doi = "10.1016/S0370-1573(99)00083-6",
    journal = "Phys. Rept.",
    volume = "323",
    pages = "183--386",
    year = "2000"
}

@article{Freund:1980xh,
  author = {Freund, Peter G. O. and Rubin, Mark A.},
  title = {Dynamics of Dimensional Reduction},
  journal = {Phys. Lett. B},
  volume = {97},
  pages = {233--235},
  year = {1980},
  doi = {10.1016/0370-2693(80)90590-0}
}

@article{Dias:2015rxy,
    author = "Dias, \'Oscar J. C. and Santos, Jorge E. and Way, Benson",
    title = "{Black holes with a single Killing vector field: black resonators}",
    eprint = "1505.04793",
    archivePrefix = "arXiv",
    primaryClass = "hep-th",
    doi = "10.1007/JHEP12(2015)171",
    journal = "JHEP",
    volume = "12",
    pages = "171",
    year = "2015"
}

@article{Hartnoll:2008kx,
    author = "Hartnoll, Sean A. and Herzog, Christopher P. and Horowitz, Gary T.",
    title = "{Holographic Superconductors}",
    eprint = "0810.1563",
    archivePrefix = "arXiv",
    primaryClass = "hep-th",
    doi = "10.1088/1126-6708/2008/12/015",
    journal = "JHEP",
    volume = "12",
    pages = "015",
    year = "2008"
}

@article{Dias:2010ma,
    author = "Dias, Oscar J. C. and Monteiro, Ricardo and Reall, Harvey S. and Santos, Jorge E.",
    title = "{A Scalar field condensation instability of rotating anti-de Sitter black holes}",
    eprint = "1007.3745",
    archivePrefix = "arXiv",
    primaryClass = "hep-th",
    doi = "10.1007/JHEP11(2010)036",
    journal = "JHEP",
    volume = "11",
    pages = "036",
    year = "2010"
}

@article{Dias:2011tj,
    author = "Dias, Oscar J. C. and Figueras, Pau and Minwalla, Shiraz and Mitra, Prahar and Monteiro, Ricardo and Santos, Jorge E.",
    title = "{Hairy black holes and solitons in global $AdS_5$}",
    eprint = "1112.4447",
    archivePrefix = "arXiv",
    primaryClass = "hep-th",
    doi = "10.1007/JHEP08(2012)117",
    journal = "JHEP",
    volume = "08",
    pages = "117",
    year = "2012"
}

@article{Horowitz:2022leb,
    author = "Horowitz, Gary T. and Kolanowski, Maciej and Santos, Jorge E.",
    title = "{A deformed IR: a new IR fixed point for four-dimensional holographic theories}",
    eprint = "2211.01385",
    archivePrefix = "arXiv",
    primaryClass = "hep-th",
    doi = "10.1007/JHEP02(2023)152",
    journal = "JHEP",
    volume = "02",
    pages = "152",
    year = "2023"
}

@article{Ishii:2018oms,
    author = "Ishii, Takaaki and Murata, Keiju",
    title = "{Black resonators and geons in AdS5}",
    eprint = "1810.11089",
    archivePrefix = "arXiv",
    primaryClass = "hep-th",
    reportNumber = "KUNS-2737, OU-HET-981",
    doi = "10.1088/1361-6382/ab1d76",
    journal = "Class. Quant. Grav.",
    volume = "36",
    number = "12",
    pages = "125011",
    year = "2019"
}

@conference{Dafermos2006,
	author = {Dafermos, M.},
	organization   = {Available at: \\
	\href{http://www-old.newton.ac.uk/webseminars/pg+ws/2006/gmx/1010/dafermos/}{http://www-old.newton.ac.uk/webseminars/pg+ws/2006/gmx/1010/dafermos/}},
	publisher = {University of Cambridge},
	title = {The Black Hole Stability problem},
	booktitle = {Talk at the Newton Institute},
	year = "2006",
}

@conference{DafermosHolzegel2006,
	author = {Dafermos, M. and Holzegel, G.},
	organization   = {Available at: \href{https://www.dpmms.cam.ac.uk/~md384/ADSinstability.pdf}{https://www.dpmms.cam .ac.uk/$\sim$md384/ADSinstability.pdf}},
	publisher = {University of Cambridge},
	title = {Dynamic instability of solitons in 4+1 dimensional gravity with negative cosmological constant},
	booktitle = {Seminar at DAMTP},
	year = "2006",
}

@article{friedrich86,
  author   = {Friedrich, H.},
  doi      = {10.1016/0393-0440(86)90004-5},
  journal  = {J. Geom. Phys.},
  pages    = {101--117},
  title    = {Existence and structure of past asymptotically simple solutions of Einstein's field equations with positive cosmological constant},
  volume   = {3},
  year     = {1986}
}

@book{Christodoulou:1993uv,
      author         = "Christodoulou, D. and Klainerman, S.",
      title          = "{The Global nonlinear stability of the Minkowski space}",
     publisher = "{Princeton Univ. Press}",
      year           = "1993",
      SLACcitation   = "%%CITATION = INSPIRE-370148;%%"
}

@article{Bizon:2011gg,
      author         = "Bizon, Piotr and Rostworowski, Andrzej",
      title          = "{On weakly turbulent instability of anti-de Sitter
                        space}",
      journal        = "Phys. Rev. Lett.",
      volume         = "107",
      year           = "2011",
      pages          = "031102",
      doi            = "10.1103/PhysRevLett.107.031102",
      eprint         = "1104.3702",
      archivePrefix  = "arXiv",
      primaryClass   = "gr-qc",
      SLACcitation   = "%%CITATION = ARXIV:1104.3702;%%"
}

@article{Dias:2011ss,
      author         = "Dias, Oscar J. C. and Horowitz, Gary T. and Santos, Jorge
                        E.",
      title          = "{Gravitational Turbulent Instability of Anti-de Sitter
                        Space}",
      journal        = "Class. Quant. Grav.",
      volume         = "29",
      year           = "2012",
      pages          = "194002",
      doi            = "10.1088/0264-9381/29/19/194002",
      eprint         = "1109.1825",
      archivePrefix  = "arXiv",
      primaryClass   = "hep-th",
      SLACcitation   = "%%CITATION = ARXIV:1109.1825;%%"
}

@article{Horowitz:2014hja,
      author         = "Horowitz, Gary T. and Santos, Jorge E.",
      title          = "{Geons and the Instability of Anti-de Sitter Spacetime}",
      journal        = "Surveys Diff. Geom.",
      volume         = "20",
      year           = "2015",
      pages          = "321-335",
      doi            = "10.4310/SDG.2015.v20.n1.a13",
      eprint         = "1408.5906",
      archivePrefix  = "arXiv",
      primaryClass   = "gr-qc",
      SLACcitation   = "%%CITATION = ARXIV:1408.5906;%%"
}

@article{Dias:2012tq,
      author         = "Dias, Oscar J. C. and Horowitz, Gary T. and Marolf, Don
                        and Santos, Jorge E.",
      title          = "{On the Nonlinear Stability of Asymptotically Anti-de
                        Sitter Solutions}",
      journal        = "Class. Quant. Grav.",
      volume         = "29",
      year           = "2012",
      pages          = "235019",
      doi            = "10.1088/0264-9381/29/23/235019",
      eprint         = "1208.5772",
      archivePrefix  = "arXiv",
      primaryClass   = "gr-qc",
      SLACcitation   = "%%CITATION = ARXIV:1208.5772;%%"
}

@article{Maliborski:2012gx,
      author         = "Maliborski, Maciej",
      title          = "{Instability of Flat Space Enclosed in a Cavity}",
      journal        = "Phys. Rev. Lett.",
      volume         = "109",
      year           = "2012",
      pages          = "221101",
      doi            = "10.1103/PhysRevLett.109.221101",
      eprint         = "1208.2934",
      archivePrefix  = "arXiv",
      primaryClass   = "gr-qc",
      SLACcitation   = "%%CITATION = ARXIV:1208.2934;%%"
}

@article{Buchel:2012uh,
      author         = "Buchel, Alex and Lehner, Luis and Liebling, Steven L.",
      title          = "{Scalar Collapse in AdS}",
      journal        = "Phys. Rev.",
      volume         = "D86",
      year           = "2012",
      pages          = "123011",
      doi            = "10.1103/PhysRevD.86.123011",
      eprint         = "1210.0890",
      archivePrefix  = "arXiv",
      primaryClass   = "gr-qc",
      reportNumber   = "UWO-TH-12-10",
      SLACcitation   = "%%CITATION = ARXIV:1210.0890;%%"
}

@article{Basu:2012gg,
      author         = "Basu, Pallab and Das, Diptarka and Das, Sumit R. and
                        Nishioka, Tatsuma",
      title          = "{Quantum Quench Across a Zero Temperature Holographic
                        Superfluid Transition}",
      journal        = "JHEP",
      volume         = "03",
      year           = "2013",
      pages          = "146",
      doi            = "10.1007/JHEP03(2013)146",
      eprint         = "1211.7076",
      archivePrefix  = "arXiv",
      primaryClass   = "hep-th",
      SLACcitation   = "%%CITATION = ARXIV:1211.7076;%%"
}

@ARTICLE{2012arXiv1212.1907G,
   author = {{Gannot}, O.},
    title = "{Quasinormal modes for Schwarzschild-AdS black holes: exponential convergence to the real axis}",
  journal = {ArXiv e-prints},
archivePrefix = "arXiv",
   eprint = {1212.1907},
 primaryClass = "math.SP",
     year = 2012,
    month = dec,
   adsurl = {http://adsabs.harvard.edu/abs/2012arXiv1212.1907G}
}

@article{Maliborski:2013jca,
      author         = "Maliborski, Maciej and Rostworowski, Andrzej",
      title          = "{Time-Periodic Solutions in an Einstein
                        AdS-Massless-Scalar-Field System}",
      journal        = "Phys. Rev. Lett.",
      volume         = "111",
      year           = "2013",
      pages          = "051102",
      doi            = "10.1103/PhysRevLett.111.051102",
      eprint         = "1303.3186",
      archivePrefix  = "arXiv",
      primaryClass   = "gr-qc",
      SLACcitation   = "%%CITATION = ARXIV:1303.3186;%%"
}

@article{Buchel:2013uba,
      author         = "Buchel, Alex and Liebling, Steven L. and Lehner, Luis",
      title          = "{Boson stars in AdS spacetime}",
      journal        = "Phys. Rev.",
      volume         = "D87",
      year           = "2013",
      number         = "12",
      pages          = "123006",
      doi            = "10.1103/PhysRevD.87.123006",
      eprint         = "1304.4166",
      archivePrefix  = "arXiv",
      primaryClass   = "gr-qc",
      reportNumber   = "UWO-TH-13-5",
      SLACcitation   = "%%CITATION = ARXIV:1304.4166;%%"
}

@article{Bizon:2013xha,
      author         = "Bizon, Piotr and Jamuna, Joanna",
      title          = "{Globally regular instability of $AdS_3$}",
      journal        = "Phys. Rev. Lett.",
      volume         = "111",
      year           = "2013",
      number         = "4",
      pages          = "041102",
      doi            = "10.1103/PhysRevLett.111.041102",
      eprint         = "1306.0317",
      archivePrefix  = "arXiv",
      primaryClass   = "gr-qc",
      SLACcitation   = "%%CITATION = ARXIV:1306.0317;%%"
}

@article{Maliborski:2013ula,
      author         = "Maliborski, Maciej and Rostworowski, Andrzej",
      title          = "{A comment on "Boson stars in AdS"}",
      year           = "2013",
      eprint         = "1307.2875",
      archivePrefix  = "arXiv",
      SLACcitation   = "%%CITATION = ARXIV:1307.2875;%%"
}

@article{Baier:2013gsa,
      author         = "Baier, R. and Stricker, S. A. and Taanila, O.",
      title          = "{Critical scalar field collapse in AdS$_{3}$: an
                        analytical approach}",
      journal        = "Class. Quant. Grav.",
      volume         = "31",
      year           = "2014",
      pages          = "025007",
      doi            = "10.1088/0264-9381/31/2/025007",
      eprint         = "1309.1629",
      archivePrefix  = "arXiv",
      primaryClass   = "gr-qc",
      reportNumber   = "BI-TP-2013-19, TUW-13-13",
      SLACcitation   = "%%CITATION = ARXIV:1309.1629;%%"
}

@article{Jalmuzna:2013rwa,
      author         = "Jamuna, Joanna",
      title          = "{Three-dimensional Gravity and Instability of AdS$_{3}$}",
      journal        = "Acta Phys. Polon.",
      volume         = "B44",
      year           = "2013",
      number         = "12",
      pages          = "2603-2620",
      doi            = "10.5506/APhysPolB.44.2603",
      eprint         = "1311.7409",
      archivePrefix  = "arXiv",
      primaryClass   = "gr-qc",
      SLACcitation   = "%%CITATION = ARXIV:1311.7409;%%"
}

@article{Fodor:2013lza,
      author         = "Fodor, Gyula and Forg\'acs, P\'eter and Grandcl\'ement,
                        Philippe",
      title          = "{Scalar field breathers on anti-de Sitter background}",
      journal        = "Phys. Rev.",
      volume         = "D89",
      year           = "2014",
      number         = "6",
      pages          = "065027",
      doi            = "10.1103/PhysRevD.89.065027",
      eprint         = "1312.7562",
      archivePrefix  = "arXiv",
      primaryClass   = "hep-th",
      SLACcitation   = "%%CITATION = ARXIV:1312.7562;%%"
}

@article{Friedrich:2014raa,
      author         = "Friedrich, Helmut",
      title          = "{On the AdS stability problem}",
      journal        = "Class. Quant. Grav.",
      volume         = "31",
      year           = "2014",
      pages          = "105001",
      doi            = "10.1088/0264-9381/31/10/105001",
      eprint         = "1401.7172",
      archivePrefix  = "arXiv",
      primaryClass   = "gr-qc",
      SLACcitation   = "%%CITATION = ARXIV:1401.7172;%%"
}

@article{Maliborski:2014rma,
      author         = "Maliborski, Maciej and Rostworowski, Andrzej",
      title          = "{What drives AdS spacetime unstable?}",
      journal        = "Phys. Rev.",
      volume         = "D89",
      year           = "2014",
      number         = "12",
      pages          = "124006",
      doi            = "10.1103/PhysRevD.89.124006",
      eprint         = "1403.5434",
      archivePrefix  = "arXiv",
      primaryClass   = "gr-qc",
      SLACcitation   = "%%CITATION = ARXIV:1403.5434;%%"
}

@article{Abajo-Arrastia:2014fma,
      author         = "Abajo-Arrastia, Javier and da Silva, Emilia and Lopez,
                        Esperanza and Mas, Javier and Serantes, Alexandre",
      title          = "{Holographic Relaxation of Finite Size Isolated Quantum
                        Systems}",
      journal        = "JHEP",
      volume         = "05",
      year           = "2014",
      pages          = "126",
      doi            = "10.1007/JHEP05(2014)126",
      eprint         = "1403.2632",
      archivePrefix  = "arXiv",
      primaryClass   = "hep-th",
      reportNumber   = "IFT-UAM-CSIC-14-012",
      SLACcitation   = "%%CITATION = ARXIV:1403.2632;%%"
}

@article{Balasubramanian:2014cja,
      author         = "Balasubramanian, Venkat and Buchel, Alex and Green,
                        Stephen R. and Lehner, Luis and Liebling, Steven L.",
      title          = "{Holographic Thermalization, Stability of Anti-de
                        Sitter Space, and the Fermi-Pasta-Ulam Paradox}",
      journal        = "Phys. Rev. Lett.",
      volume         = "113",
      year           = "2014",
      number         = "7",
      pages          = "071601",
      doi            = "10.1103/PhysRevLett.113.071601",
      eprint         = "1403.6471",
      archivePrefix  = "arXiv",
      primaryClass   = "hep-th",
      reportNumber   = "UWO-TH-14-2",
      SLACcitation   = "%%CITATION = ARXIV:1403.6471;%%"
}

@article{Bizon:2014bya,
      author         = "Bizon, Piotr and Rostworowski, Andrzej",
      title          = "{Comment on Holographic Thermalization, Stability of
                        Anti-de Sitter Space, and the Fermi-Pasta-Ulam
                        Paradox?}",
      journal        = "Phys. Rev. Lett.",
      volume         = "115",
      year           = "2015",
      number         = "4",
      pages          = "049101",
      doi            = "10.1103/PhysRevLett.115.049101",
      eprint         = "1410.2631",
      archivePrefix  = "arXiv",
      primaryClass   = "gr-qc",
      SLACcitation   = "%%CITATION = ARXIV:1410.2631;%%"
}

@article{daSilva:2014zva,
      author         = "da Silva, Emilia and Lopez, Esperanza and Mas, Javier and
                        Serantes, Alexandre",
      title          = "{Collapse and Revival in Holographic Quenches}",
      journal        = "JHEP",
      volume         = "04",
      year           = "2015",
      pages          = "038",
      doi            = "10.1007/JHEP04(2015)038",
      eprint         = "1412.6002",
      archivePrefix  = "arXiv",
      primaryClass   = "hep-th",
      SLACcitation   = "%%CITATION = ARXIV:1412.6002;%%"
}

@article{Balasubramanian:2015uua,
      author         = "Balasubramanian, Venkat and Buchel, Alex and Green,
                        Stephen R. and Lehner, Luis and Liebling, Steven L.",
      title          = "{Reply to Comment on Holographic Thermalization,
                        Stability of Anti-de Sitter Space, and the
                        Fermi-Pasta-Ulam Paradox?}",
      journal        = "Phys. Rev. Lett.",
      volume         = "115",
      year           = "2015",
      number         = "4",
      pages          = "049102",
      doi            = "10.1103/PhysRevLett.115.049102",
      eprint         = "1506.07907",
      archivePrefix  = "arXiv",
      primaryClass   = "gr-qc",
      SLACcitation   = "%%CITATION = ARXIV:1506.07907;%%"
}

@article{Craps:2014vaa,
      author         = "Craps, Ben and Evnin, Oleg and Vanhoof, Joris",
      title          = "{Renormalization group, secular term resummation and AdS
                        (in)stability}",
      journal        = "JHEP",
      volume         = "10",
      year           = "2014",
      pages          = "48",
      doi            = "10.1007/JHEP10(2014)048",
      eprint         = "1407.6273",
      archivePrefix  = "arXiv",
      primaryClass   = "gr-qc",
      SLACcitation   = "%%CITATION = ARXIV:1407.6273;%%"
}

@article{Basu:2014sia,
      author         = "Basu, Pallab and Krishnan, Chethan and Saurabh, Ayush",
      title          = "{A stochasticity threshold in holography and the
                        instability of AdS}",
      journal        = "Int. J. Mod. Phys.",
      volume         = "A30",
      year           = "2015",
      number         = "21",
      pages          = "1550128",
      doi            = "10.1142/S0217751X15501286",
      eprint         = "1408.0624",
      archivePrefix  = "arXiv",
      primaryClass   = "hep-th",
      SLACcitation   = "%%CITATION = ARXIV:1408.0624;%%"
}

@article{Deppe:2014oua,
      author         = "Deppe, Nils and Kolly, Allison and Frey, Andrew and
                        Kunstatter, Gabor",
      title          = "{Stability of AdS in Einstein Gauss Bonnet Gravity}",
      journal        = "Phys. Rev. Lett.",
      volume         = "114",
      year           = "2015",
      pages          = "071102",
      doi            = "10.1103/PhysRevLett.114.071102",
      eprint         = "1410.1869",
      archivePrefix  = "arXiv",
      primaryClass   = "hep-th",
      SLACcitation   = "%%CITATION = ARXIV:1410.1869;%%"
}

@article{Dimitrakopoulos:2014ada,
      author         = "Dimitrakopoulos, Fotios V. and Freivogel, Ben and
                        Lippert, Matthew and Yang, I-Sheng",
      title          = "{Position space analysis of the AdS (in)stability
                        problem}",
      journal        = "JHEP",
      volume         = "08",
      year           = "2015",
      pages          = "077",
      doi            = "10.1007/JHEP08(2015)077",
      eprint         = "1410.1880",
      archivePrefix  = "arXiv",
      primaryClass   = "hep-th",
      SLACcitation   = "%%CITATION = ARXIV:1410.1880;%%"
}

@article{Buchel:2014xwa,
      author         = "Buchel, Alex and Green, Stephen R. and Lehner, Luis and
                        Liebling, Steve L.",
      title          = "{Conserved quantities and dual turbulent cascades in
                        anti-de Sitter spacetime}",
      journal        = "Phys. Rev.",
      volume         = "D91",
      year           = "2015",
      number         = "6",
      pages          = "064026",
      doi            = "10.1103/PhysRevD.91.064026",
      eprint         = "1412.4761",
      archivePrefix  = "arXiv",
      primaryClass   = "gr-qc",
      SLACcitation   = "%%CITATION = ARXIV:1412.4761;%%"
}

@article{Craps:2014jwa,
      author         = "Craps, Ben and Evnin, Oleg and Vanhoof, Joris",
      title          = "{Renormalization, averaging, conservation laws and AdS
                        (in)stability}",
      journal        = "JHEP",
      volume         = "01",
      year           = "2015",
      pages          = "108",
      doi            = "10.1007/JHEP01(2015)108",
      eprint         = "1412.3249",
      archivePrefix  = "arXiv",
      primaryClass   = "gr-qc",
      SLACcitation   = "%%CITATION = ARXIV:1412.3249;%%"
}

@article{Basu:2015efa,
      author         = "Basu, Pallab and Krishnan, Chethan and Bala Subramanian,
                        P. N.",
      title          = "{AdS (In)stability: Lessons From The Scalar Field}",
      journal        = "Phys. Lett.",
      volume         = "B746",
      year           = "2015",
      pages          = "261-265",
      doi            = "10.1016/j.physletb.2015.05.009",
      eprint         = "1501.07499",
      archivePrefix  = "arXiv",
      primaryClass   = "hep-th",
      SLACcitation   = "%%CITATION = ARXIV:1501.07499;%%"
}

@article{Yang:2015jha,
      author         = "Yang, I-Sheng",
      title          = "{Missing top of the AdS resonance structure}",
      journal        = "Phys. Rev.",
      volume         = "D91",
      year           = "2015",
      number         = "6",
      pages          = "065011",
      doi            = "10.1103/PhysRevD.91.065011",
      eprint         = "1501.00998",
      archivePrefix  = "arXiv",
      primaryClass   = "hep-th",
      SLACcitation   = "%%CITATION = ARXIV:1501.00998;%%"
}

@article{Okawa:2015xma,
      author         = "Okawa, Hirotada and Lopes, Jorge C. and Cardoso, Vitor",
      title          = "{Collapse of massive fields in anti-de Sitter spacetime}",
      year           = "2015",
      eprint         = "1504.05203",
      archivePrefix  = "arXiv",
      primaryClass   = "gr-qc",
      SLACcitation   = "%%CITATION = ARXIV:1504.05203;%%"
}

@article{Bizon:2015pfa,
      author         = "Bizon, Piotr and Maliborski, Maciej and Rostworowski,
                        Andrzej",
      title          = "{Resonant Dynamics and the Instability of Anti-de
                        Sitter Spacetime}",
      journal        = "Phys. Rev. Lett.",
      volume         = "115",
      year           = "2015",
      number         = "8",
      pages          = "081103",
      doi            = "10.1103/PhysRevLett.115.081103",
      eprint         = "1506.03519",
      archivePrefix  = "arXiv",
      primaryClass   = "gr-qc",
      SLACcitation   = "%%CITATION = ARXIV:1506.03519;%%"
}

@article{Dimitrakopoulos:2015pwa,
      author         = "Dimitrakopoulos, Fotios and Yang, I-Sheng",
      title          = "{Conditionally extended validity of perturbation theory:
                        Persistence of AdS stability islands}",
      journal        = "Phys. Rev.",
      volume         = "D92",
      year           = "2015",
      number         = "8",
      pages          = "083013",
      doi            = "10.1103/PhysRevD.92.083013",
      eprint         = "1507.02684",
      archivePrefix  = "arXiv",
      primaryClass   = "hep-th",
      SLACcitation   = "%%CITATION = ARXIV:1507.02684;%%"
}

@article{Green:2015dsa,
      author         = "Green, Stephen R. and Maillard, Antoine and Lehner, Luis
                        and Liebling, Steven L.",
      title          = "{Islands of stability and recurrence times in AdS}",
      year           = "2015",
      eprint         = "1507.08261",
      archivePrefix  = "arXiv",
      primaryClass   = "gr-qc",
      SLACcitation   = "%%CITATION = ARXIV:1507.08261;%%"
}

@article{Deppe:2015qsa,
      author         = "Deppe, Nils and Frey, Andrew R.",
      title          = "{Classes of Stable Initial Data for Massless and Massive
                        Scalars in Anti-de Sitter Spacetime}",
      year           = "2015",
      eprint         = "1508.02709",
      archivePrefix  = "arXiv",
      primaryClass   = "hep-th",
      SLACcitation   = "%%CITATION = ARXIV:1508.02709;%%"
}

@article{Craps:2015iia,
      author         = "Craps, Ben and Evnin, Oleg and Vanhoof, Joris",
      title          = "{Ultraviolet asymptotics and singular dynamics of AdS
                        perturbations}",
      year           = "2015",
      eprint         = "1508.04943",
      archivePrefix  = "arXiv",
      primaryClass   = "gr-qc",
      SLACcitation   = "%%CITATION = ARXIV:1508.04943;%%"
}

@article{Craps:2015xya,
      author         = "Craps, Ben and Evnin, Oleg and Jai-akson, Puttarak and
                        Vanhoof, Joris",
      title          = "{Ultraviolet asymptotics for quasiperiodic AdS4
                        perturbations}",
      year           = "2015",
      eprint         = "1508.05474",
      archivePrefix  = "arXiv",
      primaryClass   = "gr-qc",
      SLACcitation   = "%%CITATION = ARXIV:1508.05474;%%"
}

@article{Evnin:2015gma,
      author         = "Evnin, Oleg and Krishnan, Chethan",
      title          = "{A Hidden Symmetry of AdS Resonances}",
      journal        = "Phys. Rev.",
      volume         = "D91",
      year           = "2015",
      number         = "12",
      pages          = "126010",
      doi            = "10.1103/PhysRevD.91.126010",
      eprint         = "1502.03749",
      archivePrefix  = "arXiv",
      primaryClass   = "hep-th",
      SLACcitation   = "%%CITATION = ARXIV:1502.03749;%%"
}

@article{Menon:2015oda,
      author         = "Menon, Dhanya S. and Suneeta, Vardarajan",
      title          = "{Necessary conditions for an AdS-type instability}",
      year           = "2015",
      eprint         = "1509.00232",
      archivePrefix  = "arXiv",
      primaryClass   = "gr-qc",
      SLACcitation   = "%%CITATION = ARXIV:1509.00232;%%"
}

@article{Jalmuzna:2015hoa,
      author         = "Jalmuzna, Joanna and Gundlach, Carsten and Chmaj,
                        Tadeusz",
      title          = "{Scalar field critical collapse in 2+1 dimensions}",
      journal        = "Phys. Rev.",
      volume         = "D92",
      year           = "2015",
      number         = "12",
      pages          = "124044",
      doi            = "10.1103/PhysRevD.92.124044",
      eprint         = "1510.02592",
      archivePrefix  = "arXiv",
      primaryClass   = "gr-qc",
      SLACcitation   = "%%CITATION = ARXIV:1510.02592;%%"
}

@article{Evnin:2015wyi,
      author         = "Evnin, Oleg and Nivesvivat, Rongvoram",
      title          = "{AdS perturbations, isometries, selection rules and the
                        Higgs oscillator}",
      journal        = "JHEP",
      volume         = "01",
      year           = "2016",
      pages          = "151",
      doi            = "10.1007/JHEP01(2016)151",
      eprint         = "1512.00349",
      archivePrefix  = "arXiv",
      primaryClass   = "hep-th",
      SLACcitation   = "%%CITATION = ARXIV:1512.00349;%%"
}

@article{Freivogel:2015wib,
      author         = "Freivogel, Ben and Yang, I-Sheng",
      title          = "{Coherent Cascade: Collapsing Solutions in Global AdS}",
      year           = "2015",
      eprint         = "1512.04383",
      archivePrefix  = "arXiv",
      primaryClass   = "hep-th",
      SLACcitation   = "%%CITATION = ARXIV:1512.04383;%%"
}

@article{Fodor:2015eia,
      author         = "Fodor, Gyula and Forg\'acs, Peter and Grandcl\'ement,
                        Philippe",
      title          = "{Self-gravitating scalar breathers with negative
                        cosmological constant}",
      journal        = "Phys. Rev.",
      volume         = "D92",
      year           = "2015",
      number         = "2",
      pages          = "025036",
      doi            = "10.1103/PhysRevD.92.025036",
      eprint         = "1503.07746",
      archivePrefix  = "arXiv",
      primaryClass   = "gr-qc",
      SLACcitation   = "%%CITATION = ARXIV:1503.07746;%%"
}

@article{Dias:2016ewl,
      author         = "Dias, O and Santos, Jorge E.",
      title          = "{AdS nonlinear instability: moving beyond spherical
                        symmetry}",
      journal        = "Class. Quant. Grav.",
      volume         = "33",
      year           = "2016",
      number         = "23",
      pages          = "23LT01",
      doi            = "10.1088/0264-9381/33/23/23LT01",
      eprint         = "1602.03890",
      archivePrefix  = "arXiv",
      primaryClass   = "hep-th",
      SLACcitation   = "%%CITATION = ARXIV:1602.03890;%%"
}

@article{Rostworowski:2016isb,
      author         = "Rostworowski, Andrzej",
      title          = "{Comment on "AdS nonlinear instability: moving beyond
                        spherical symmetry" [Class. Quantum Grav. 33 23LT01
                        (2016)]}",
      year           = "2016",
      eprint         = "1612.00042",
      archivePrefix  = "arXiv",
      primaryClass   = "hep-th",
      SLACcitation   = "%%CITATION = ARXIV:1612.00042;%%"
}

@conference{Bizon2014,
	author = {Piotr, B.},
	organization   = {Available at: \\
	\href{http://physics.princeton.edu/strings2014/slides/Bizon.pdf}{http://physics.princeton.edu/strings2014/slides/Bizon.pdf}},
	publisher = {Princeton University},
	title = {Gravitational turbulent instability of AdS$_5$},
	booktitle = {Talk at Strings},
	year = "2014",
}

@article{Evnin:2016mjx,
      author         = "Evnin, Oleg and Jai-akson, Puttarak",
      title          = "{Detailed ultraviolet asymptotics for AdS scalar field
                        perturbations}",
      journal        = "JHEP",
      volume         = "04",
      year           = "2016",
      pages          = "054",
      doi            = "10.1007/JHEP04(2016)054",
      eprint         = "1602.05859",
      archivePrefix  = "arXiv",
      primaryClass   = "hep-th",
      SLACcitation   = "%%CITATION = ARXIV:1602.05859;%%"
}

@article{Deppe:2016gur,
      author         = "Deppe, Nils",
      title          = "{On the stability of anti-de Sitter spacetime}",
      year           = "2016",
      eprint         = "1606.02712",
      archivePrefix  = "arXiv",
      primaryClass   = "gr-qc",
      SLACcitation   = "%%CITATION = ARXIV:1606.02712;%%"
}

@article{Jalmuzna:2017mpa,
      author         = "Jalmuzna, Joanna and Gundlach, Carsten",
      title          = "{Critical collapse of a rotating scalar field in $2+1$
                        dimensions}",
      journal        = "Phys. Rev.",
      volume         = "D95",
      year           = "2017",
      number         = "8",
      pages          = "084001",
      doi            = "10.1103/PhysRevD.95.084001",
      eprint         = "1702.04601",
      archivePrefix  = "arXiv",
      primaryClass   = "gr-qc",
      SLACcitation   = "%%CITATION = ARXIV:1702.04601;%%"
}

@article{Moschidis:2017lcr,
      author         = "Moschidis, Georgios",
      title          = "{The Einstein--null dust system in spherical symmetry
                        with an inner mirror: structure of the maximal development
                        and Cauchy stability}",
      year           = "2017",
      eprint         = "1704.08685",
      archivePrefix  = "arXiv",
      primaryClass   = "gr-qc",
      SLACcitation   = "%%CITATION = ARXIV:1704.08685;%%"
}

@article{Moschidis:2017llu,
      author         = "Moschidis, Georgios",
      title          = "{A proof of the instability of AdS for the Einstein--null
                        dust system with an inner mirror}",
      year           = "2017",
      eprint         = "1704.08681",
      archivePrefix  = "arXiv",
      primaryClass   = "gr-qc",
      SLACcitation   = "%%CITATION = ARXIV:1704.08681;%%"
}

@article{Dias:2017tjg,
      author         = "Dias, Oscar J. C. and Santos, Jorge E.",
      title          = "{AdS nonlinear instability: breaking spherical and axial
                        symmetries}",
      year           = "2017",
      eprint         = "1705.03065",
      archivePrefix  = "arXiv",
      primaryClass   = "hep-th",
      SLACcitation   = "%%CITATION = ARXIV:1705.03065;%%"
}

@article{Dimitrakopoulos:2016tss,
      author         = "Dimitrakopoulos, Fotios V. and Freivogel, Ben and
                        Pedraza, Juan F. and Yang, I-Sheng",
      title          = "{Gauge dependence of the AdS instability problem}",
      journal        = "Phys. Rev.",
      volume         = "D94",
      year           = "2016",
      number         = "12",
      pages          = "124008",
      doi            = "10.1103/PhysRevD.94.124008",
      eprint         = "1607.08094",
      archivePrefix  = "arXiv",
      primaryClass   = "hep-th",
      SLACcitation   = "%%CITATION = ARXIV:1607.08094;%%"
}

@article{Dimitrakopoulos:2016euh,
      author         = "Dimitrakopoulos, Fotios V. and Freivogel, Ben and
                        Pedraza, Juan F.",
      title          = "{Fast and Slow Coherent Cascades in Anti-de Sitter
                        Spacetime}",
      year           = "2016",
      eprint         = "1612.04758",
      archivePrefix  = "arXiv",
      primaryClass   = "hep-th",
      SLACcitation   = "%%CITATION = ARXIV:1612.04758;%%"
}

@article{Rostworowski:2017tcx,
      author         = "Rostworowski, Andrzej",
      title          = "{Higher order perturbations of Anti-de Sitter space and
                        time-periodic solutions of vacuum Einstein equations}",
      year           = "2017",
      eprint         = "1701.07804",
      archivePrefix  = "arXiv",
      primaryClass   = "gr-qc",
      reportNumber   = "CERN-TH-2017-023",
      SLACcitation   = "%%CITATION = ARXIV:1701.07804;%%"
}

@article{Martinon:2017uyo,
      author         = "Martinon, Gr\'egoire and Fodor, Gyula and Grandcl\'ement,
                        Philippe and Forg\`acs, Peter",
      title          = "{Gravitational geons in asymptotically anti-de Sitter
                        spacetimes}",
      year           = "2017",
      eprint         = "1701.09100",
      archivePrefix  = "arXiv",
      primaryClass   = "gr-qc",
      SLACcitation   = "%%CITATION = ARXIV:1701.09100;%%"
}

@article{Choptuik:2017cyd,
    author = "Choptuik, Matthew W. and Dias, \'Oscar J. C. and Santos, Jorge E. and Way, Benson",
    title = "{Collapse and Nonlinear Instability of AdS Space with Angular Momentum}",
    eprint = "1706.06101",
    archivePrefix = "arXiv",
    primaryClass = "hep-th",
    doi = "10.1103/PhysRevLett.119.191104",
    journal = "Phys. Rev. Lett.",
    volume = "119",
    number = "19",
    pages = "191104",
    year = "2017"
}

@article{Bantilan:2017kok,
      author         = "Bantilan, Hans and Figueras, Pau and Kunesch, Markus and
                        Romatschke, Paul",
      title          = "{Non-Spherically Symmetric Collapse in Asymptotically AdS
                        Spacetimes}",
      year           = "2017",
      eprint         = "1706.04199",
      archivePrefix  = "arXiv",
      primaryClass   = "hep-th",
      SLACcitation   = "%%CITATION = ARXIV:1706.04199;%%"
}

@article{Choptuik:2018ptp,
    author = "Choptuik, Matthew and Santos, Jorge E. and Way, Benson",
    title = "{Charting Islands of Stability with Multioscillators in anti\textendash{}de Sitter space}",
    eprint = "1803.02830",
    archivePrefix = "arXiv",
    primaryClass = "hep-th",
    doi = "10.1103/PhysRevLett.121.021103",
    journal = "Phys. Rev. Lett.",
    volume = "121",
    number = "2",
    pages = "021103",
    year = "2018"
}

@article{Michalogiorgakis:2006jc,
    author = "Michalogiorgakis, Georgios and Pufu, Silviu S.",
    title = "{Low-lying gravitational modes in the scalar sector of the global AdS(4) black hole}",
    eprint = "hep-th/0612065",
    archivePrefix = "arXiv",
    reportNumber = "PUPT-2218",
    doi = "10.1088/1126-6708/2007/02/023",
    journal = "JHEP",
    volume = "02",
    pages = "023",
    year = "2007"
}

@article{Hawking:1999dp,
    author = "Hawking, S. W. and Reall, H. S.",
    title = "{Charged and rotating AdS black holes and their CFT duals}",
    eprint = "hep-th/9908109",
    archivePrefix = "arXiv",
    reportNumber = "DAMTP-R-99-108",
    doi = "10.1103/PhysRevD.61.024014",
    journal = "Phys. Rev. D",
    volume = "61",
    pages = "024014",
    year = "2000"
}

@misc{DafermosRodnianskiPC,
  author       = {I. Rodnianski and M. Dafermos},
  title        = {Private communication},
  year         = {year},  % Optional: add the year if known
  note         = {Private communication}
}

@misc{KehleMoschidis,
  author       = {C. Kehle and G. Moschidis},
  title        = {In preparation},
  year         = {year},  % Optional: add the year if known
  note         = {Private communication}
}

@article{Horowitz:1999jd,
    author = "Horowitz, Gary T. and Hubeny, Veronika E.",
    title = "{Quasinormal modes of AdS black holes and the approach to thermal equilibrium}",
    eprint = "hep-th/9909056",
    archivePrefix = "arXiv",
    reportNumber = "NSF-ITP-99-70",
    doi = "10.1103/PhysRevD.62.024027",
    journal = "Phys. Rev. D",
    volume = "62",
    pages = "024027",
    year = "2000"
}

@article{Cardoso:2001bb,
    author = "Cardoso, Vitor and Lemos, Jose P. S.",
    title = "{Quasinormal modes of Schwarzschild anti-de Sitter black holes: Electromagnetic and gravitational perturbations}",
    eprint = "gr-qc/0105103",
    archivePrefix = "arXiv",
    reportNumber = "DF-IST-4-2001",
    doi = "10.1103/PhysRevD.64.084017",
    journal = "Phys. Rev. D",
    volume = "64",
    pages = "084017",
    year = "2001"
}

@article{Figueras:2023ihz,
    author = "Figueras, Pau and Rossi, Lorenzo",
    title = "{Non-linear instability of slowly rotating Kerr-AdS black holes}",
    eprint = "2311.14167",
    archivePrefix = "arXiv",
    primaryClass = "hep-th",
    month = "11",
    year = "2023"
}

@article{Holzegel:2011uu,
    author = "Holzegel, Gustav and Smulevici, Jacques",
    title = "{Decay properties of Klein-Gordon fields on Kerr-AdS spacetimes}",
    eprint = "1110.6794",
    archivePrefix = "arXiv",
    primaryClass = "gr-qc",
    doi = "10.1002/cpa.21470",
    journal = "Commun. Pure Appl. Math.",
    volume = "66",
    pages = "1751--1802",
    year = "2013"
}

@article{bondi1960,
  author  = {Bondi, H.},
  title   = {Gravitational Waves in General Relativity},
  journal = {Nature},
  year    = {1960},
  volume  = {186},
  number  = {4724},
  pages   = {535},
  doi     = {10.1038/186535a0} 
}

@article{Balasubramanian_2014,
   title={Losing forward momentum holographically},
   volume={31},
   ISSN={1361-6382},
   DOI={10.1088/0264-9381/31/12/125010},
   number={12},
   journal={Classical and Quantum Gravity},
   publisher={IOP Publishing},
   author={Balasubramanian, K. and Herzog, C. P.},
   year={2014}, pages={125010} }

@article{Skenderis_2006,
   title={Kaluza-Klein holography},
   volume={2006},
   ISSN={1029-8479},
   DOI={10.1088/1126-6708/2006/05/057},
   number={05},
   journal={Journal of High Energy Physics},
   publisher={Springer Science and Business Media LLC},
   author={Skenderis, K. and Taylor, M.},
   year={2006},
   pages={057–057} }

@article{Balasubramanian_1999,
   title={A Stress Tensor for Anti-de Sitter Gravity},
   volume={208},
   ISSN={1432-0916},
   DOI={10.1007/s002200050764},
   number={2},
   journal={Communications in Mathematical Physics},
   publisher={Springer Science and Business Media LLC},
   author={Balasubramanian, V. and Kraus, P.},
   year={1999},
   pages={413–428} }

@article{Hawking_1996,
   title={The gravitational Hamiltonian, action, entropy and surface terms},
   volume={13},
   ISSN={1361-6382},
   DOI={10.1088/0264-9381/13/6/017},
   number={6},
   journal={Classical and Quantum Gravity},
   publisher={IOP Publishing},
   author={Hawking, S. W. and Horowitz, G. T.},
   year={1996},
   pages={1487–1498} }

@book{Trefethen,
author    = {Trefethen, L. N.},
title     = {Spectral Methods in MATLAB},
publisher = {Society for Industrial and Applied Mathematics},
year      = {2000},
doi       = {10.1137/1.9780898719598}
}

@book{CanutoI,
  title        = {Spectral Methods: Fundamentals in Single Domains},
  author       = {Canuto, C. and Youssuff Hussaini, M. and Quarteroni, A. and Zang, T. A. },
  series       = {Scientific Computation},
  edition      = {1},
  publisher    = {Springer Berlin, Heidelberg},
  year         = {2006},
  doi          = {10.1007/978-3-540-30726-6}
}

@book{CanutoII,
  title        = {Spectral Methods: Evolution to Complex Geometries and Applications to Fluid Dynamics},
  author       = {Canuto, C. and Quarteroni, A. and Yousuff Hussaini, M. and Zang, T. A.},
  series       = {Scientific Computation},
  edition      = {1},
  publisher    = {Springer Berlin, Heidelberg},
  year         = {2007},
  doi          = {10.1007/978-3-540-30728-0}
}

@article{HolzegelI,
    author = "Graf, O. and Holzegel, G.",
    title = "{Linear Stability of Schwarzschild-Anti-de Sitter spacetimes I: The system of gravitational perturbations}",
    eprint = "2408.02251",
    archivePrefix = "arXiv",
    primaryClass = "gr-qc",
    year = "2024"
}

@article{HolzegelII,
    author = "Graf, O. and Holzegel, G.",
    title = "{Linear Stability of Schwarzschild-Anti-de Sitter spacetimes II: Logarithmic decay of solutions to the Teukolsky system}",
    eprint = "2408.02252",
    archivePrefix = "arXiv",
    primaryClass = "gr-qc",
    year = "2024"
}

@article{Cardoso_2001,
   title={Quasinormal modes of Schwarzschild–anti-de Sitter black holes: Electromagnetic and gravitational perturbations},
   volume={64},
   ISSN={1089-4918},
   DOI={10.1103/physrevd.64.084017},
   number={8},
   journal={Physical Review D},
   publisher={American Physical Society (APS)},
   author={Cardoso, V. and Lemos, J. P. S.},
   year={2001}}

@article{Cardoso_2003,
   title={Quasinormal frequencies of Schwarzschild black holes in anti–de Sitter spacetimes: A complete study of the overtone asymptotic behavior},
   volume={68},
   ISSN={1089-4918},
   DOI={10.1103/physrevd.68.044024},
   number={4},
   journal={Physical Review D},
   publisher={American Physical Society (APS)},
   author={Cardoso, V. and Konoplya, R. and Lemos, J. P. S.},
   year={2003} }

@article{Buchel_2015,
doi = {10.1088/0264-9381/32/14/145003},
year = {2015},
publisher = {IOP Publishing},
volume = {32},
number = {14},
pages = {145003},
author = {Buchel, A. and Lehner, L.},
title = {Small black holes in AdS5 × S5},
journal = {Classical and Quantum Gravity}
}

@article{Hubeny_2002,
doi = {10.1088/1126-6708/2002/05/027},
year = {2002},
publisher = {},
volume = {2002},
number = {05},
pages = {027},
author = {Hubeny, V. E.  and Rangamani, M.},
title = {Unstable Horizons},
journal = {Journal of High Energy Physics}
}

@article{Dias_2012,
   title={On the nonlinear stability of asymptotically anti-de Sitter solutions},
   volume={29},
   ISSN={1361-6382},
   DOI={10.1088/0264-9381/29/23/235019},
   number={23},
   journal={Classical and Quantum Gravity},
   publisher={IOP Publishing},
   author={Santos, J. E. and Dias, O. J. C. and Horowitz, G. T. and Marolf, D.},
   year={2012},
   pages={235019} }

@article{Berrut2004BarycentricLI,
  title={Barycentric Lagrange Interpolation},
  author={Berrut, J. P. and Trefethen, L. N.},
  journal={SIAM Rev.},
  year={2004},
  volume={46},
  pages={501-517},
  DOI={10.1137/S0036144502417715}
}

@book{Boyd_1989, 
title={Chebyshev \& Fourier Spectral Methods}, 
ISSN={0176-5035}, 
DOI={10.1007/978-3-642-83876-7}, 
journal={Lecture Notes in Engineering}, 
publisher={Springer Berlin Heidelberg}, 
author={Boyd, J. P.}, 
year={1989} }

@article{Warnick_2014,
   title={On Quasinormal Modes of Asymptotically Anti-de Sitter Black Holes},
   volume={333},
   ISSN={1432-0916},
   DOI={10.1007/s00220-014-2171-1},
   number={2},
   journal={Communications in Mathematical Physics},
   publisher={Springer Science and Business Media LLC},
   author={Warnick, C. M.},
   year={2014},
   month=sep, pages={959–1035} }

@inproceedings{Moschidis_2023,
    author = {Moschidis, G. and Kehle, C. },
    booktitle= {Spectral Theory and Mathematical Relativity},
    address = {Erwin Schrodinger International Institute for Mathematics and Physics},
    title = {Weak turbulence on Schwarzschild-AdS spacetime},
    year = {2023},
    url = {https://www.dpmms.cam.ac.uk/~rbdt2/NAGR/NAGR_17_Moschidis.pdf}

}

@article{KNOLL2004357,
title = {Jacobian-free Newton–Krylov methods: a survey of approaches and applications},
journal = {Journal of Computational Physics},
volume = {193},
number = {2},
pages = {357-397},
year = {2004},
issn = {0021-9991},
doi = {https://doi.org/10.1016/j.jcp.2003.08.010},
author = {D.A. Knoll and D.E. Keyes},

}

@article{GMRES,
author = {Saad, Youcef and Schultz, Martin H.},
title = {GMRES: A Generalized Minimal Residual Algorithm for Solving Nonsymmetric Linear Systems},
journal = {SIAM Journal on Scientific and Statistical Computing},
volume = {7},
number = {3},
pages = {856-869},
year = {1986},
doi = {10.1137/0907058}

}

\end{document}